\documentclass[twocolumn]{aastex63}
\usepackage{multirow}

\usepackage{multirow}
\usepackage{soul}
\usepackage{upgreek}
\hypersetup{linkcolor=blue,citecolor=blue,filecolor=magenta,urlcolor=cyan}

\submitjournal{AAS Journal}

\shorttitle{Logos-Zoe}
\shortauthors{Thirouin et al.}

\begin{document}

\title{Logos-Zoe: A contact binary triple system in the trans-neptunian belt.}

\correspondingauthor{Audrey Thirouin}
\email{thirouin@lowell.edu}

\author[0000-0002-1506-4248]{Audrey Thirouin}
\affiliation{Lowell Observatory, 1400 W Mars Hill Road, Flagstaff, Arizona, 86001, USA}

\author[0000-0002-6013-9384]{Keith S. Noll}
\affiliation{NASA Goddard Space Flight Center, 8800 Greenbelt Road, Greenbelt, Maryland, 20771, USA}

\author[0000-0002-8296-6540]{William M. Grundy}
\affiliation{Lowell Observatory, 1400 W Mars Hill Road, Flagstaff, Arizona, 86001, USA}

\author[0000-0003-3145-8682]{Scott S. Sheppard}
\affiliation{Earth and Planets Laboratory, Carnegie Institution for Science, \\ 5241 Broad Branch Rd. NW, Washington, District of Columbia, 20015, USA.}

\author{Felicity Escarzaga}
\affiliation{Lowell Observatory, 1400 W Mars Hill Road, Flagstaff, Arizona, 86001, USA}

\author{Brian Donnelly}
\affiliation{School of Informatics, Computing, and Cyber Systems, Northern Arizona University, P.O. Box 5693, Flagstaff, AZ, 86011, USA}



\begin{abstract}
The trans-neptunian object (58534) 1997~CQ$_{29}$ (a.k.a. Logos) is a resolved wide binary in the dynamically Cold Classical population. With \emph{Hubble Space Telescope} resolved observations where the primary Logos is well separated from its secondary Zoe it can be established that Logos has a time-variable brightness. Logos' brightness varied by several tenths of a magnitude over a short timescale of hours while the brightness variability of Zoe was on a longer timescale. New unresolved ground-based observations obtained with the \emph{Lowell Discovery Telescope} and the \emph{Magellan-Baade} telescope confirm at least one highly variable component in this system. With our ground-based observations and photometric constraints from space-based observations, we suggest that the primary Logos is likely a close/contact binary whose rotational period is 17.43$\pm$0.06~h for a lightcurve amplitude of 0.70$\pm$0.07~mag while Zoe is potentially a (very) slow rotator with an unknown shape. Using the \texttt{Candela} software, we model the Logos-Zoe system and predict its upcoming mutual events season using rotational, physical, and mutual orbit parameters derived in this work or already published. Zoe's shape and rotational period are still uncertain, so we consider various options to better understand Zoe. The upcoming mutual event season for Logos-Zoe starts in 2026 and will last for four years with up to two events per year. Observations of these mutual events will allow us to significantly improve the physical and rotational properties of both Logos and Zoe.   

\end{abstract}



\section{Introduction} 
\label{sec:intro}

The trans-neptunian object (58534) 1997~CQ$_{29}$ Logos was discovered on February 4$^{\textnormal{th}}$, 1997 with the 2.2~m University of Hawaii telescope in Mauna Kea Observatory \citep{Chen1997} while its secondary Zoe was one of the earliest-discovered satellites with the \textit{Hubble Space Telescope (HST)} on November 17$^{\textnormal{th}}$, 2001 \citep{Noll2002IAU, Noll2002}. With a semimajor$\footnote{\url{https://minorplanetcenter.net/db_search/show_object?object_id=58534&commit=Show}}$ axis of 45.020~au, an eccentricity of 0.121, and an inclination of 2.90$^\circ$, the system belongs to the dynamically Cold Classical population and displays the characteristic ultra-red surface color inferring an in-situ formation \citep{Tegler1998, Barucci2000, Davies2000, JewittLuu2001, Boehnhardt2001, GilHuttonLicandro2001, McBride2003, Barucci2008}.  

Based on \textit{HST} images, the combined absolute magnitude of the system is H$_{V}^{LZ}$=7.28$\pm$0.05~mag and the magnitude difference between Logos and Zoe is $\Delta _{mag}^{L-Z}$$\sim$0.45~mag \citep{Grundy2011}. At first glance, the Logos-Zoe system is a typical Cold Classical binary with two similar brightness components with maximum effective diameters of $\sim$80~km for Logos, and $\sim$66~km for Zoe \citep{Noll2004}. The two components orbit a common center of mass with an orbital period of P$_{orb}$=309.87$\pm$0.22~days in a moderately eccentric, e$_{orb}$=0.5463$\pm$0.0079, high inclination, i$_{orb}$=95.43$\pm$0.47$^{\circ}$ retrograde orbit, and the system will undergo a mutual event season with the satellite passing behind or in front of the primary in $\sim$2027 \citep{Grundy2011}. The system mass is M$_{sys}$=(0.458$\pm$0.007)$\times$10$^{18}$ kg, and so if the Logos-Zoe albedo is $\sim$0.15, similar to that of Arrokoth and other Cold Classicals, the system's density would be $\sim$300~kg m$^{-3}$ if both components are spheres \citep{Brucker2009, Grundy2011, Lacerda2014HSO, Stern2019}. However, \citet{Noll2004, Grundy2005} suggested that the system's albedo can be as high as 0.39$\pm$0.17 considering a density between 0.5 and 2~g/cm$^{3}$ and spherical components. The dynamically Cold Classical population has a large fraction of wide binaries compared to the other trans-neptunian sub-populations with a fraction of 22$^{+10}_{-5}$~\% and $\sim$10 to 25~\% of Cold Classicals are likely contact binaries \citep{Noll2020, ThirouinSheppard2019, Noll2008}. Cold Classical binaries tend to be (nearly-) equal-sized wide binaries, and all ``large'' Cold Classical with an absolute magnitude less than 6.5-7~mag are wide binaries \citep{Noll2020}. Therefore, the Logos-Zoe system with two similar brightness components inferring a (nearly-) equal size binary and an ultra-red surface is a fairly typical Cold Classical object compared to the rest of the sub-population's members. The Logos-Zoe's orbital period is fairly long at about 310~days, which is not odd as orbital periods can range from 5.5 days to 6590~days \citep{Grundy2019}. However, the interesting feature of the mutual orbit is that the orbit is retrograde while most of the binary/multiple Cold Classical trans-neptunian systems tend to be prograde \citep{Grundy2019}.

The mutual events season prediction and other derived parameters (e.g., density,  albedo) assume that the system has two spherical components, which is the by-default assumption in such studies \citep{Grundy2011}. However, several pieces of evidence point to the fact that at least one of the components is not spherical (see Section~\ref{sec:lit}), and thus most estimates and predictions for this system are likely off. Physical characteristics, such as density and rotational properties, are keys to understanding planetesimal and planet formation during the early stages of the Solar System. Therefore, by taking advantage of the upcoming mutual events season of the Logos-Zoe system, we have an incredible opportunity to understand this system thanks to the synergy between modeling, ground- and space-based observations. We note that such mutual events season happen only every 151~years for Logos-Zoe.  

Therefore, we design a study whose main focus is to constrain the system's characteristics with new ground-based observations in support of already published and/or public space-based observations as well as lightcurve modeling improving our understanding of the Logos-Zoe system. In Section~\ref{sec:lit}, we will summarize the evidence obtained over the past two decades inferring that the system is more complicated than initially thought. Sections~\ref{sec:obs} and \ref{sec:interpretation} will present the newly obtained ground-based dataset and its interpretation. Finally, we will model the system to predict its upcoming mutual events season in Section~\ref{sec:mutualeventseason}. 


\section{Ground- and space-based observations of Logos-Zoe: summary of published and public datasets} 
\label{sec:lit}  

Over the past two decades, Logos-Zoe has been observed with ground- and space-based telescopes. The \textit{HST} allows \textit{resolved observations} as the two components are separated and thus the photometry of each component can be evaluated individually. While with ground-based observations, it is impossible to identify each component even under excellent weather conditions, thus, we measure the photometry of the pair with the \textit{unresolved observations}. 

\subsection{Resolved observations with \textit{Hubble Space Telescope}}

\startlongtable
\begin{deluxetable*}{ c|c|c|c|c|c }
\tablecaption{\label{HST}  Published photometric observations of the Logos-Zoe system with \textit{HST} are listed. Magnitudes have been converted to the V- and I-bands \citep{Noll2004, Benecchi2009}. Values shown in italics (this work) are calculated using the component V$-$I colors measured on 23 June 2004 with appropriate error propagation. Note that the component identity on 18 June 2002 when Logos is fainter than Zoe has been verified by orbit fitting \citep{Grundy2011}. Using the 2004 HST dataset, \citet{Benecchi2009} reported that the colors of Logos and Zoe are, respectively: (V-I)$_{Logos}$=1.05$\pm$0.08~mag and (V-I)$_{Zoe}$=1.09$\pm$0.18~mag. Based on \citet{Hainaut2012} database, the average color of the Cold Classical sub-population is 1.23$\pm$0.16~mag, therefore, Logos and Zoe, as well as the rest of the Cold Classicals, have similar surface colors. }
\tablewidth{0pt}
\tablehead{   & \multicolumn{2}{c|}{V-band [mag]} & \multicolumn{2}{c|}{I-band [mag]} &  References      \\  \hline
Observing Date    & Logos        & Zoe          & Logos        & Zoe          &                            }
\startdata
17 November 2001  & 23.78$\pm$0.13        & 24.02$\pm$0.22        & 22.54$\pm$0.09        & 23.06$\pm$0.19        & \multirow{4}{*}{\citet{Noll2002, Noll2004}} \\ \cline{1-5}
18 June 2002      & \it{24.27$\pm$0.09}   & \it{24.12$\pm$0.18}   & 23.42$\pm$0.04        & 23.03$\pm$0.03        &                                   \\ \cline{1-5}
30 June 2002      & \it{23.68$\pm$0.09}   & \it{24.05$\pm$0.18}   & 22.63$\pm$0.03        & 22.96$\pm$0.04        &                                   \\ \cline{1-5}
12 July 2002      & \it{23.74$\pm$0.09}   & \it{24.15$\pm$0.18}   & 22.69$\pm$0.03        & 23.06$\pm$0.04        &                                   \\ \hline
23 June 2004      & 23.52$\pm$0.06   & 23.85$\pm$0.06   & 22.47$\pm$0.05        & 23.51$\pm$0.18        & \citet{Benecchi2009}              \\ \hline
17 December 2007  & 24.14$\pm$0.04        & 24.60$\pm$0.04        & \it{23.09$\pm$0.09}   &  \it{23.51$\pm$0.18}  & \citet{Grundy2011}                \\ 
\hline 
 \enddata
\end{deluxetable*}

Logos has been the target of several \textit{HST} proposals$\footnote{\textit{HST} programs: \#9060 (PI: K.S. Noll), \#9386 (PI: K.S. Noll), \#9746 (PI: J.-L. Margot), \#9585 (PI: K.S. Noll), \#11178 (PI: W.M. Grundy).}$ first discovering its binary nature and then deriving its mutual orbit \citep{Noll2002, Noll2004, Margot2005, Grundy2011}. Early on, \citet{Noll2002} noticed that the system shows evidence of large amplitude variability, but no constraints on the rotational periods of the components could be derived due to the sparse observation cadence (Table~\ref{HST}). Although the multi-epoch observations used to determine the binary mutual orbit are sparse, the $\sim$0.8~mag change in Logos from 18$^{\textnormal{th}}$ to 30$^{\textnormal{th}}$ June 2002 shows that this component is highly variable and must be rotating asynchronously. Logos’ brightness difference from November 2001 to June 2002 was even higher at nearly 0.9~mag (Table~\ref{HST}). Zoe did not vary significantly in the 2001-2002 epochs, but observations taken in December 2007 show it to be nearly 0.6~mag fainter and by about 0.5~mag fainter in 2004 (Table~\ref{HST}). 

In conclusion, potentially both components of the binary show large amplitude variabilities. However, the timescales are not well determined from the existing \textit{HST} photometry alone as the data are too sparse \citep{Noll2008}.  
\subsection{Unresolved observations with ground-based telescopes}
   
The near-infrared and visible colors of several Centaurs and trans-neptunian objects (TNOs) were derived by \citet{Boehnhardt2001} with facilities in Chile and Spain. Logos$\footnote{Zoe was not discovered yet at the time of \citet{Boehnhardt2001} publication.}$ was one of their targets for JHKs and VRI photometry (as well as spectroscopy). Based on V-band measurements obtained on May 13$^{\textnormal{th}}$ and 14$^{\textnormal{th}}$ 1999, \citet{Boehnhardt2001} noticed that the object's brightness varied by about 0.38~mag. They suggested that this variability is due to the object's rotation inferring a non-spherical shape or/and albedo spot(s). \citet{McBride2003} noticed the inconsistent color measurements obtained by \citet{JewittLuu2001, Boehnhardt2001} and proposed that this discrepancy could be attributed to a large lightcurve amplitude and/or the binary nature of the system. 
 
On March 18$^{\textnormal{th}}$ 2017, \citet{ThirouinSheppard2019} obtained three images of Logos-Zoe with the \textit{Lowell Discovery Telescope}. They published a lower limit for the brightness variation of 0.50~mag over approximately 1~h. As stated in \citet{ThirouinSheppard2019}, these observations were to confirm the large variability of the system already noticed nearly two decades before with ground- and space-based observations. The feasibility of future observations with this telescope/instrument was also tested as Logos-Zoe is faint with a visual magnitude of $\sim$23.5~mag and a large photometric variability.
 
\section{New ground-based observations: lightcurve, interpretation and basic modeling} 
\label{sec:obs} 

\subsection{New Observations}

 To derive Logos-Zoe's ground-based lightcurve, we obtained observing time from May 2019 to April 2023 with the \textit{Lowell Discovery Telescope (LDT)} and the \textit{Magellan-Baade} telescope (Table~\ref{Obs_log}).

 The \textit{LDT} is a 4~m telescope in Arizona, and is equipped with the Large Monolithic Imager (LMI) which has a 12.5$\arcmin$$\times$12.5$\arcmin$ field of view for a pixel scale of 0.12$\arcsec$/pixel \citep{Levine2012}. The \textit{Magellan-Baade} is a 6~m facility at Las Campanas Observatory in Chile with a  wide-field imager called IMACS which has a 27.4$\arcmin$ diameter field for a 0.20$\arcsec$/pixel scale \citep{Dressler2011}. To ensure a high enough signal-to-noise ratio for Logos-Zoe, observations with broad-band filters --VR at \textit{LDT} and WB4800–7800 (WB for short in Table~\ref{Obs_log}) at \textit{Magellan-Baade}-- and exposure times between 350~s and 800~s were obtained were obtained. 
 
For the newly obtained set of ground-based observations, we followed the data calibration and reduction summarized in \citet{Thirouin2014, Thirouin2010}. The interpretation of the lightcurve is based on its morphology and amplitude following \citet{ThirouinSheppard2024} (see Section~\ref{sec:LCtext}).

\startlongtable
\begin{deluxetable}{cccccc}
 \tabletypesize{\scriptsize}
\tablecaption{\label{Obs_log} Logos-Zoe ground-based observing log for our 2019-2023 campaign. Heliocentric (r$_h$) and geocentric ($\Delta$) distances and phase angles ($\alpha$) are computed by the Minor Planet Center. \\  }
\tablewidth{0pt}
\tablehead{  UT  &   r$_h$ &  $\Delta$ & $\alpha$   & Filter & Telescope   \\
       Observing Date       &  [AU]  &  [AU]  &  [$^{\circ}$]   & &    }
\startdata
4 May 2019      &42.552  & 43.341   &0.8 & VR &   LDT \\
19 May 2020        &42.837&  43.453 &1.1 & VR &   LDT \\
22 December 2020      &  43.639 & 43.517 &1.3 & VR &   LDT \\
8 February 2021        &42.877 & 43.531 &1.0 & VR &   LDT \\
13 April 2021        &42.586 & 43.550   &0.4 & VR &   LDT \\
14 May 2021        &42.860 & 43.560  &1.0 & VR &   LDT \\
26 Mar 2022        &42.660 & 43.654  &0.1 & VR &   LDT \\
8 April 2022        &42.671 & 43.658&0.2 & WB &   Magellan \\
9 April 2022        &42.674 & 43.658 &0.2 & WB&   Magellan \\
29 April 2022        & 42.792  &43.664  &0.7 &WB &   Magellan \\
30 April 2022        & 42.801 & 43.665   &0.7 &WB &   Magellan \\
27 March 2023        &42.771  & 43.765&0.1 & VR &   LDT \\
21 April 2023        &42.834&  43.773&0.5 & VR &   LDT \\
22 April 2023        &42.839 & 43.773 &0.5 & VR &   LDT \\
27 April 2023        &42.875  &43.775&0.6 & VR &   LDT \\
 \hline
\hline
\enddata
\end{deluxetable}

\subsection{Lightcurve and Interpretation}
\label{sec:LCtext}

As we are dealing with a binary system, several periodicities have to be taken into account: the orbital period of the two components in the system (P$_{orb}$), (2) the rotational period of the primary (P$_{Logos}$), and (3) the rotational period of the satellite (P$_{Zoe}$). The orbital period P$_{orb}$ is well known \citep{Grundy2011}, but P$_{Logos}$ and P$_{Zoe}$ are both unknown. Based on the space-based observations, it seems that Logos has a short-term variability of hours while Zoe's variability can be in the range of days to months (Section~\ref{sec:lit}). Therefore, it is likely that the Logos-Zoe system is in an asynchronous state with P$_{orb}$$\neq$P$_{Logos}$$\neq$P$_{Zoe}$. So far, only a handful of trans-neptunian binaries have been classified as asynchronous systems: Haumea-Hi'iaka \citep{Hastings2016}, 2003~QY$_{90}$ \citep{KernElliot2006}, and Logos-Zoe (this work). However, as suggested by \citet{Grundy2012}, systems with a non (nearly-) circular mutual orbit are likely in an asynchronous state. Based on the binary/multiple systems with known mutual orbit, we can estimate that up to 35 systems can be in an asynchronous state. We note that because only a handful of binaries/multiples have known rotational lightcurves and even fewer satellites have a known rotational period, the tidal effects, tidal locking, and despinning effects remain highly unexplored in the trans-Neptunian belt. Even if the synchronization time can be estimated and is generally longer than the age of the Solar System for most of the binaries, the tidal effects may have already started to slow down the rotation of the system's components \citep{Gladman1996, Thirouin2014}.  

  \begin{figure}
  \includegraphics[width=9cm, angle=0]{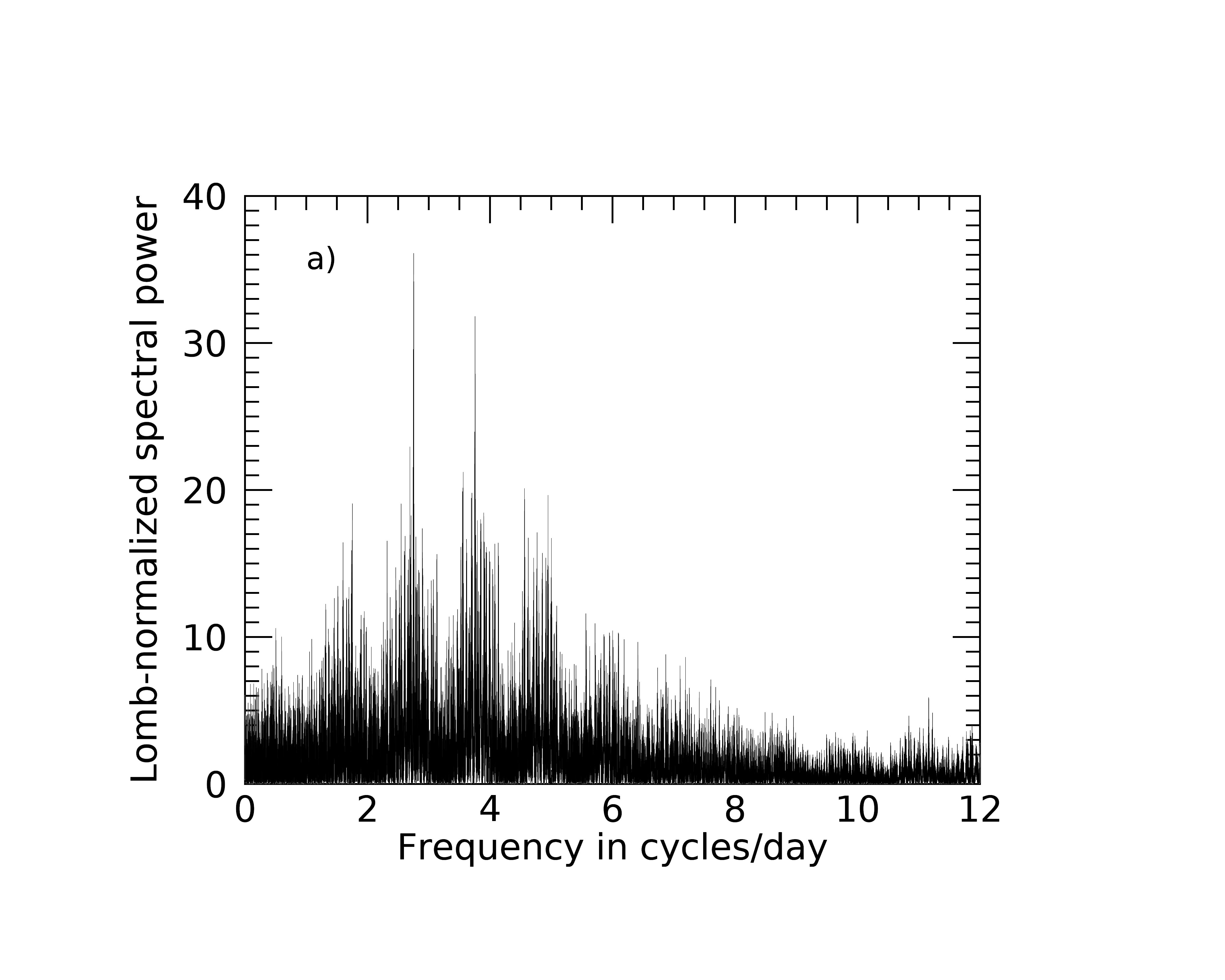}
  \includegraphics[width=9cm, angle=0]{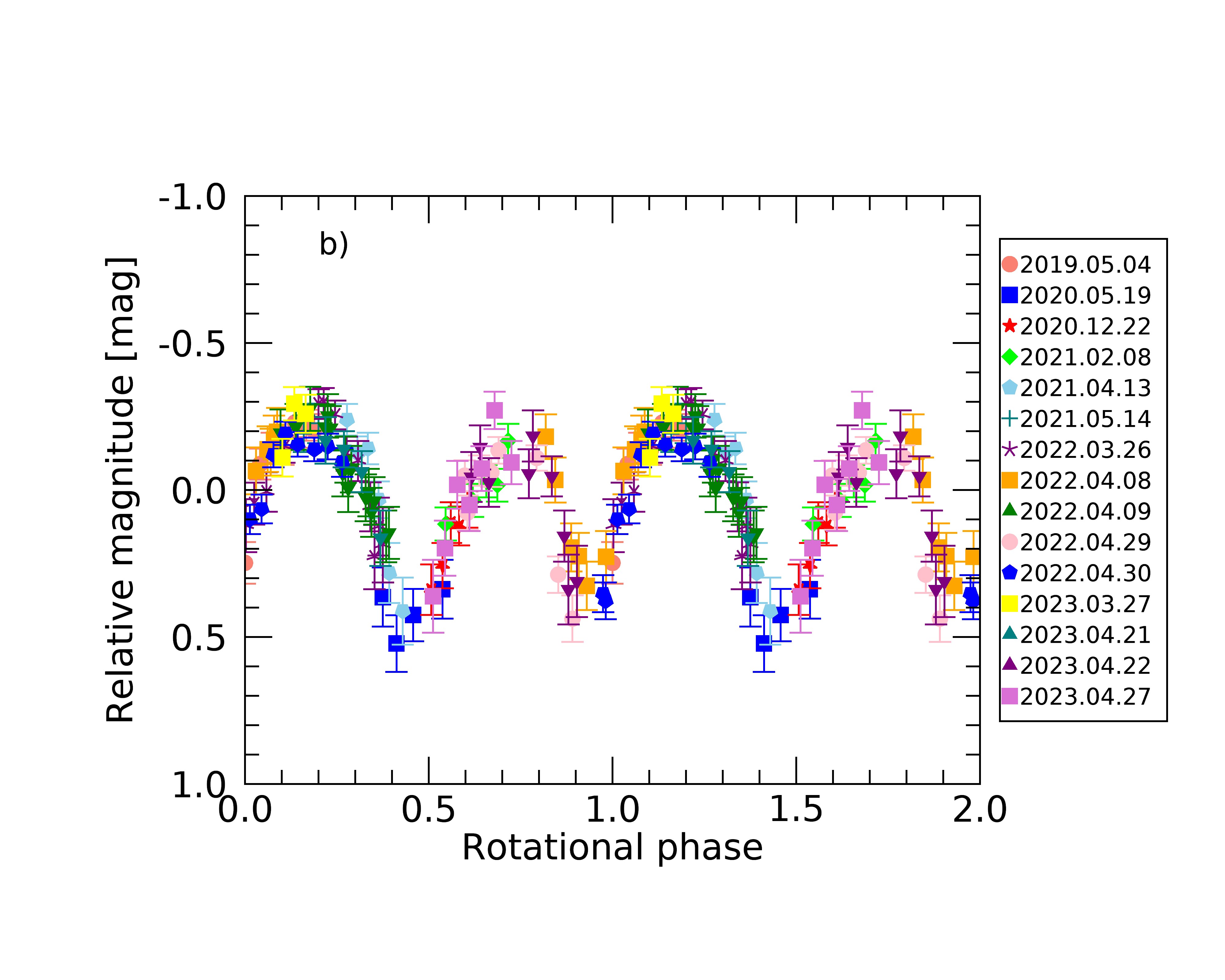}
  \includegraphics[width=9cm, angle=0]{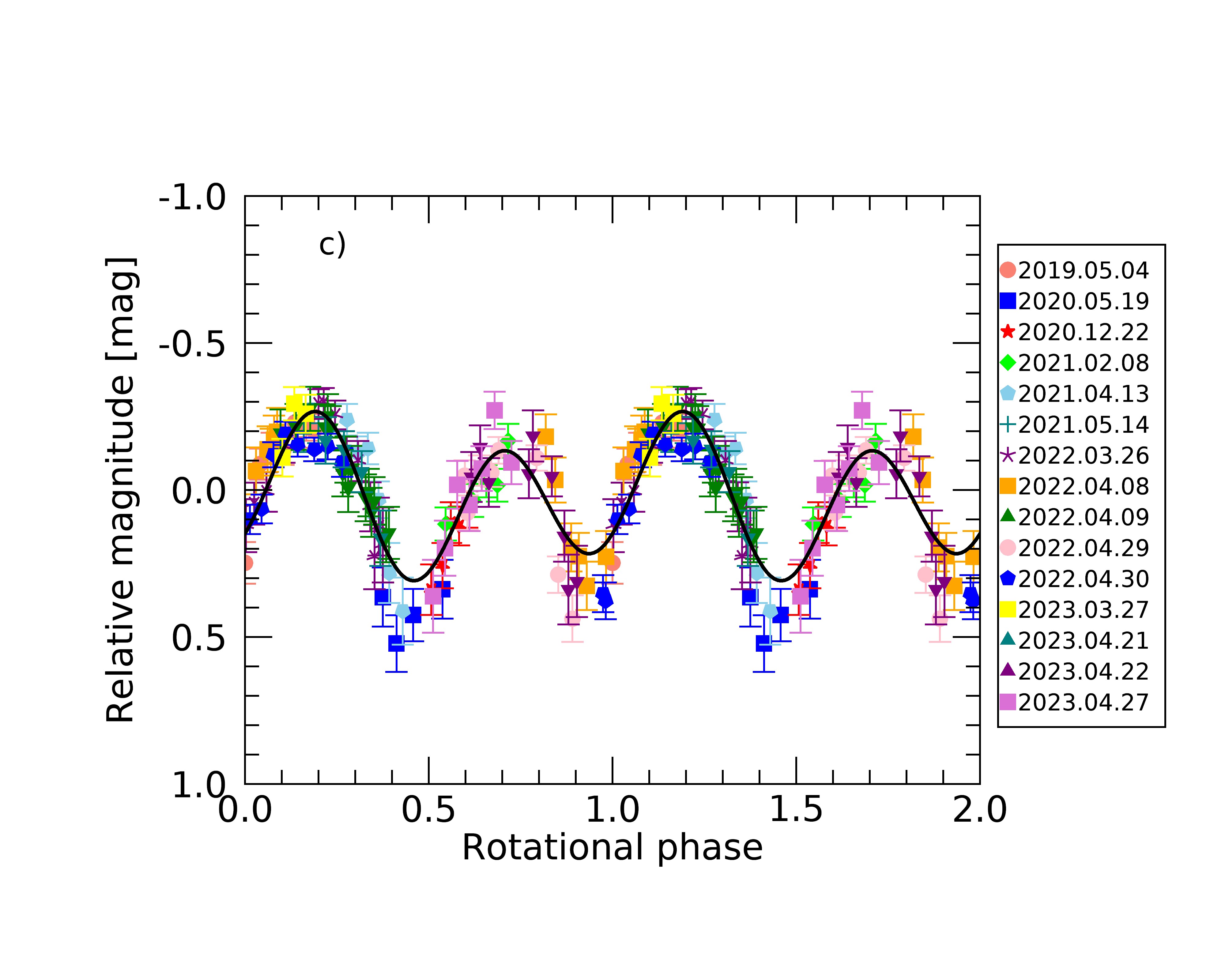}
\caption{The most significant peak of the Lomb periodogram is located at 2.75~cycles/day (Plot a)). However, because of the lightcurve's asymmetry and large amplitude, we select the double-peaked periodicity of 17.43$\pm$0.06~h (Plot b)). The overplotted Fourier series fit (black continuous line) is unable to match the data inferring that the lightcurve is not sinusoidal (Plot c)). The lightcurve is plotted over two rotations (i.e., rotational phase between 0 and 2) for an improved visualization of the cyclical brightness variation.}
\label{fig:LC_LZ}
\end{figure} 

The new ground-based photometric dataset was obtained over 5 years. We investigated every year separately and there has been no amplitude change over the years due to changes in the system's geometry. Therefore, we treated all the data as one unique dataset. 

Using the system's photometry obtained from 2019 to 2023, we searched for periodicity with the Lomb periodogram technique \citep{Lomb1976}. The Lomb periodogram peak with the highest confidence level is at 2.75~cycles/day (Plot a) of Figure~\ref{fig:LC_LZ}). A frequency of f$^{single}$=2.75~cycles/day is equivalent to a period of P$^{single}$=24/f$^{single}$=8.72~h. Because of the asymmetric lightcurve and large amplitude, the double-peaked lightcurve (with a rotational period of P$^{double}$=2$\times$P$^{single}$) is the optimal solution \citep{ThirouinSheppard2024}. Due to the large variability over a short timescale of hours, we interpret that the lightcurve in Plot b) of Figure~\ref{fig:LC_LZ} is Logos'. Therefore, we suggest that P$^{double}_{Logos}$=17.43$\pm$0.06~h for an amplitude of $\Delta m$$_{Logos}$=0.70$\pm$0.07~mag. 

Typically, the lightcurve amplitude of a spheroidal object is less than 0.15~mag, thus, we can discard that Logos is a spheroidal object due to its extreme variability \citep{ThirouinSheppard2024}. The second-order Fourier series fit (Plot c) of Figure~\ref{fig:LC_LZ}) is unable to replicate our observational dataset with a $\chi^{2}$=1.46. So, because Logos' lightcurve is not sinusoidal, we can also discard an ellipsoidal shape \citep{ThirouinSheppard2024}. Therefore, we propose that Logos is a likely close/contact binary$\footnote{We note that the meaning of close/contact binary comprises objects with a bilobed shape, two objects touching in one point, and objects separated by a few hundred kilometers \citep{ThirouinSheppard2024}. In the literature, these variations tend to be classified as contact binary. }$ based on its U-/V-shaped lightcurve and large amplitude. 

The full amplitude is less than the 0.9~mag limit used by \citet{Leone1984, Weidenschilling1980} to distinguish a contact binary from a highly elongated object, but Logos' brightness variation was likely larger at nearly 0.9~mag between November 2001 and June 2002 based on \textit{HST} images. As pointed out by \citet{Lacerda2011, ThirouinSheppard2024}, if an object is not observed equator-on, the lightcurve will have a lower amplitude and so the 0.9~mag limit will not be reached. Future observations at different epochs may show how Logos' amplitude is changing over time from different viewing geometries, possibly allowing us to determine its pole orientation. For a complete study, Logos will be modeled as a contact binary and an elongated ellipsoid in Section~\ref{sec:interpretation} and we will demonstrate that the elongated ellipsoid is not a good match to our dataset.  

An interesting feature of Logos' lightcurve is its asymmetry. The first maximum at a rotational phase of 0.2 is taller by $\sim$0.07~mag compared to the second maximum at a rotational phase of 0.7. There may also be a slight asymmetry in the timing of the peaks. These asymmetries are probably due to albedo mark(s) or to an irregular shape. For asteroids, \citet{Degewij1979} inferred that the typical albedo variations on small body surfaces are between 4 and 10~\% and it has been shown that these variations create asymmetric lightcurves \citep{JewittSheppard2002, Lacerda2008}.  
 
We also investigated our dataset for other periodicities that we can attribute to Zoe as the entire \textit{HST} dataset suggests a potential long-term variability for the satellite. However, no potential periodicities (on short- and long-term) have been retrieved with a high enough confidence level to infer Zoe's rotation. We can propose several reasons why the rotational period of Zoe was not retrieved in our dataset: (1) Zoe can have an extremely long rotational period, and/or (2) Zoe has a very low lightcurve amplitude, or (3) Zoe is in a tumbling state. 

As no lightcurve information has been retrieved for Zoe, we are unable to constrain its shape and rotational properties with our newly obtained dataset. Therefore, we have to rely on the extracted pieces of information from the space-based observations. If we do not take into account the \textit{HST} data from 2007, we can interpret that Zoe is a spheroidal object or a pole-on oriented object due to the limited object's variability over the different observational sets but its rotation remains unconstrained as it can be from hours to months. However, Zoe is $\sim$0.6~mag fainter in 2007 and about 0.5~mag fainter in 2004 compared to the 2001-2002 epoch, suggesting that Zoe can be an elongated body with a slow rotation on the timescale of days to months. If Zoe is a triaxial TNO in hydrostatic equilibrium with a full lightcurve amplitude of $\Delta m$$_{Zoe}$=0.6~mag, then its axis ratios are b/a=0.58 and c/a=0.42 assuming an equatorial view$\footnote{Because \citet{Sheppard2004phd} demonstrated that the average viewing angle is $\xi$=60$^\circ$, in this case, the axis ratios would be b/a=0.51 and c/a=0.39. }$ with $\xi$=90$^\circ$ \citep{Chandrasekhar1987}. Zoe's amplitude can be even larger than the 0.6~mag estimate which is based on two \textit{HST} epochs obtained nearly 6~years apart. With potentially an amplitude greater than 0.6~mag, we have to consider the case of Zoe being a contact binary \citep{ThirouinSheppard2024}. This scenario would be an extreme case with a contact binary orbiting a contact binary. But as the first \textit{Lucy} flyby of a small body showed us that the main belt asteroid Dinkinesh has a contact binary as a companion, we will also consider this option in Section~\ref{sec:interpretation} and Section~\ref{sec:mutualeventseason} \citep{Levison2024}. To present a complete study but also consider all potential options, we will assume different sets of characteristics for Zoe (see Section~\ref{sec:mutualeventseason}).
 
\subsection{Basic modeling}

We attribute the large amplitude double-peaked $\sim$17~hour lightcurve in Figure~\ref{fig:LC_LZ} to Logos and we suggest that Logos is a likely close/contact binary. Based on this interpretation, we can use the lightcurve for basic modeling to extract pieces of information about Logos. Following, we will name the two components of Logos as Logos~A and Logos~B.  

By studying the sequences of equilibrium for binaries with mass ratioa between 0.01 and 1, \citet{Leone1984} developed the network of Roche sequences in the plane lightcurve amplitude-rotational (see Figures 2 and 3 in \citet{Leone1984}). Because of the localization of the Logos-Zoe system on the network of sequences, the density limits are fixed to 1 and 5~g/cm$^{3}$. In fact, emulating \citet{Leone1984}, we deduced that the mass ratio of Logos is between q$_{min}$$\sim$0.7 and q$_{max}$$\sim$0.8 for a density between $\rho$$_{min}$$\sim$1~g/cm$^{3}$ and $\rho$$_{max}$$\sim$5~g/cm$^{3}$. In Table~\ref{tab:Leone}, we summarize the derived axis ratios of Logos~A ((b/a)$_{A}$ and (c/a)$_{A}$) and Logos~B ((b/a)$_{B}$ and (c/a)$_{B}$) as well as the separation$\footnote{Possible values for the parameter D$_{AB}$ are between 0 and 1 and is defined as D$_{AB}$=(a$_{A}$+a$_{B}$)/d$_{AB}$ where a$_{A}$ and a$_{B}$ are the longest axis of each component and d$_{AB}$ is the distance between the object A and the object B. If D$_{AB}$=1, object A and object B are in contact, but if D$_{AB}$ is less than 1, the two objects are not touching \citep{Leone1984}. }$ between the two (D$_{AB}$ is unitless and d$_{AB}$ is in km) for the q$_{min}$ and q$_{max}$ options. 
We emphasize that a density of 5~g/cm$^{3}$ is highly unlikely for a TNO in the size range of Logos-Zoe in the trans-neptunian belt \citep{Grundy2012, Vilenius2014, Noll2020} but, this upper value is set by the network of sequences derived by \citet{Leone1984}. Therefore, even if it is not a realistic value for a TNO, it is still a mathematically feasible solution. 

As there is no direct measurement of the albedo of the system, we used several values. Albedos of 0.04 and 0.20 are the by-default values, but we also considered an albedo of 0.15 which is the typical albedo for Cold Classical TNOs \citep{Lacerda2014HSO, Brucker2009}. However, as pointed out by \citet{Noll2004, Grundy2005}, the albedo can be significantly higher at 0.39$\pm$0.17, but they assumed that Logos and Zoe were spheres. Using these albedo options, we listed in Table~\ref{tab:Leone}, the sizes of the axes for each component as well as the separation between Logos~A and Logos~B. For our calculations, we use the absolute magnitude estimated by the Minor Planet Center. We emphasize that aside from the uncertainty of this absolute magnitude estimate, there is also the contribution of Zoe that we are not taking into account. Therefore, even if these estimates are approximate, we can use them as a starting point for the more complex modeling presented in Section~\ref{sec:interpretation}.

\begin{deluxetable*}{lllllllllllll}
\tablecaption{\label{tab:Leone} Characteristics of Logos~A and Logos~B based on \citet{Leone1984} modeling and assuming that Logos is a contact binary. Logos-Zoe absolute magnitude (H$^{LZ}_{MPC}$) is from the Minor Planet Center (MPC). Several albedos are used.    \\  }
\tablewidth{0pt}
\tablehead{   \multicolumn{1}{|c|}{(b/a)$_{A}$}                   & \multicolumn{1}{c|}{(c/a)$_{A}$} & \multicolumn{1}{c|}{(b/a)$_{B}$}                  & \multicolumn{1}{c|}{(c/a)$_{B}$}                   &  \multicolumn{1}{c|}{D$_{AB}$ }  &  \multicolumn{1}{|l|}{Albedo}  & \multicolumn{1}{l|}{a$_{A}$ [km]} & \multicolumn{1}{l|}{b$_{A}$ [km]} & \multicolumn{1}{l|}{c$_{A}$ [km]} & \multicolumn{1}{l|}{a$_{B}$ [km]} & \multicolumn{1}{l|}{b$_{B}$ [km]} & \multicolumn{1}{l|}{c$_{B}$ [km]} & \multicolumn{1}{l|}{d$_{AB}$ [km]}  }
\startdata
\multicolumn{13}{|c|}{q$_{min}$=0.7, $\rho$$_{min}$=1~g/cm$^{3}$, H$^{LZ}_{MPC}$=6.87~mag}                                                                                                                                                                                                                                                                                                                                                                                                     \\ \hline
\multicolumn{1}{|c|}{\multirow{4}{*}{0.94}} & \multicolumn{1}{c|}{\multirow{4}{*}{0.90}} & \multicolumn{1}{c|}{\multirow{4}{*}{0.91}} & \multicolumn{1}{c|}{\multirow{4}{*}{0.87}} & \multicolumn{1}{c|}{\multirow{4}{*}{0.56}} & \multicolumn{1}{c|}{0.04} & \multicolumn{1}{c|}{80}  & \multicolumn{1}{c|}{75}  & \multicolumn{1}{c|}{72}  & \multicolumn{1}{c|}{72}   & \multicolumn{1}{c|}{66}   & \multicolumn{1}{c|}{63}   
& \multicolumn{1}{c|}{272}   \\ \cline{6-13} 
\multicolumn{1}{|c|}{}                      & \multicolumn{1}{c|}{}                      & \multicolumn{1}{c|}{}                      & \multicolumn{1}{c|}{}                      & \multicolumn{1}{c|}{}   & \multicolumn{1}{c|}{0.15}    & \multicolumn{1}{c|}{41}  & \multicolumn{1}{c|}{39}  & \multicolumn{1}{c|}{37}  & \multicolumn{1}{c|}{37}   & \multicolumn{1}{c|}{34}   & \multicolumn{1}{c|}{33}   & \multicolumn{1}{c|}{140}  \\ \cline{6-13} 
\multicolumn{1}{|c|}{}                      & \multicolumn{1}{c|}{}                      & \multicolumn{1}{c|}{}                      & \multicolumn{1}{c|}{}                      & \multicolumn{1}{c|}{}         & \multicolumn{1}{c|}{0.20}              & \multicolumn{1}{c|}{36}  & \multicolumn{1}{c|}{33}  & \multicolumn{1}{c|}{32}  & \multicolumn{1}{c|}{32}   & \multicolumn{1}{c|}{29}   & \multicolumn{1}{c|}{28}   & \multicolumn{1}{c|}{121}  \\ \cline{6-13}
\multicolumn{1}{|c|}{}                      & \multicolumn{1}{c|}{}                      & \multicolumn{1}{c|}{}                      & \multicolumn{1}{c|}{}                      & \multicolumn{1}{c|}{}                    & \multicolumn{1}{c|}{0.39$\pm$0.17}   & \multicolumn{1}{c|}{25$^{+9}_{-4}$}  & \multicolumn{1}{c|}{ 24$^{+8}_{-4}$ }  & \multicolumn{1}{c|}{ 23$^{+8}_{-4}$ }  & \multicolumn{1}{c|}{ 23$^{+8}_{-4}$ }   & \multicolumn{1}{c|}{ 21$^{+7}_{-3}$ }   & \multicolumn{1}{c|}{ 20$^{+7}_{-3}$ }   & \multicolumn{1}{c|}{87$^{+29}_{-14}$  }  \\ \hline
\multicolumn{13}{|c|}{q$_{max}$=0.8, $\rho$$_{max}$=5~g/cm$^{3}$, H$^{LZ}_{MPC}$=6.87~mag}                                                                                                                                                                                                                                                                                                                                                                                                    \\ \hline
\multicolumn{1}{|c|}{\multirow{4}{*}{0.99}} & \multicolumn{1}{c|}{\multirow{4}{*}{0.98}} & \multicolumn{1}{c|}{\multirow{4}{*}{0.98}}  & \multicolumn{1}{c|}{\multirow{4}{*}{0.98}} & \multicolumn{1}{c|}{\multirow{4}{*}{0.31}} & \multicolumn{1}{c|}{0.04} & \multicolumn{1}{c|}{74}  & \multicolumn{1}{c|}{73}  & \multicolumn{1}{c|}{72}  & \multicolumn{1}{c|}{69}   & \multicolumn{1}{c|}{67}   & \multicolumn{1}{c|}{67}   & \multicolumn{1}{c|}{459}   \\ \cline{6-13} 
\multicolumn{1}{|c|}{}                      & \multicolumn{1}{c|}{}                      & \multicolumn{1}{c|}{}                      & \multicolumn{1}{c|}{}                      & \multicolumn{1}{c|}{}                   & \multicolumn{1}{c|}{0.15}    & \multicolumn{1}{c|}{38}  & \multicolumn{1}{c|}{38}  & \multicolumn{1}{c|}{37}  & \multicolumn{1}{c|}{35}   & \multicolumn{1}{c|}{35}   & \multicolumn{1}{c|}{35}   & \multicolumn{1}{c|}{237}  \\ \cline{6-13} 
\multicolumn{1}{|c|}{}                      & \multicolumn{1}{c|}{}                      & \multicolumn{1}{c|}{}                      & \multicolumn{1}{c|}{}                      & \multicolumn{1}{c|}{}               & \multicolumn{1}{c|}{0.20}       & \multicolumn{1}{c|}{33}  & \multicolumn{1}{c|}{33}  & \multicolumn{1}{c|}{32}  & \multicolumn{1}{c|}{31}   & \multicolumn{1}{c|}{30}   & \multicolumn{1}{c|}{30}   & \multicolumn{1}{c|}{205}   \\ \cline{6-13} 
\multicolumn{1}{|c|}{}                      & \multicolumn{1}{c|}{}                      & \multicolumn{1}{c|}{}                      & \multicolumn{1}{c|}{}                      & \multicolumn{1}{c|}{}                & \multicolumn{1}{c|}{0.39$\pm$0.17}      & \multicolumn{1}{c|}{24$^{+7}_{-4}$}  & \multicolumn{1}{c|}{ 23$^{+8}_{-3}$ }  & \multicolumn{1}{c|}{ 23$^{+8}_{-4}$ }  & \multicolumn{1}{c|}{ 22$^{+7}_{-4}$ }   & \multicolumn{1}{c|}{ 22$^{+7}_{-4}$ }   & \multicolumn{1}{c|}{ 22$^{+7}_{-4}$ }   & \multicolumn{1}{c|}{147$^{+49}_{-24}$  }   \\ \hline
\enddata
\end{deluxetable*}


\section{Lightcurve modeling with \texttt{Candela}}
\label{sec:interpretation}

The \texttt{Candela} software is built upon the \texttt{licht} software presented by \citet{Benecchi2021}.

\subsection{What are the \texttt{licht} and \texttt{Candela} software packages?}

The \texttt{licht} software was developed by Capstone students from Northern Arizona University’s School of Informatics, Computing, and Cyber Systems over three academic years \citep{Benecchi2021}. The main goal of \texttt{licht} was to create 3D renders of binary systems and then to use the ray-tracing technique to produce the system's lightcurve. One of the main limitations of the \texttt{licht} software was the shape implementations, as the software was only able to handle spheres. 

Using the \texttt{licht} software as a base, its highly upgraded version named \texttt{Candela} provides additional functions and features. Following, we present the basic characteristics and features of \texttt{Candela} but its full description as well as presentation of several test cases will be the topic of a different publication. 

The \texttt{Candela} software is c++ based and uses fully tested and open-source libraries$\footnote{The main library used by \texttt{Candela} is Eigen which is available at \url{https://eigen.tuxfamily.org/index.php?title=Main_Page}}$. Several observational and physical inputs are required: 
\begin{enumerate}
\item \ul{Ephemeris of the system and the Sun at a specified time, time-step, and location}. The ephemeris files for the system and the Sun are generated using the Horizons System$\footnote{\url{https://ssd.jpl.nasa.gov/horizons/}}$ at the observer location as well as for a time and time-step specified by the user \citep{Giorgini1996, Giorgini2001}. Both ephemeris files (one for the system and one for the Sun) have several columns with the Julian Date, the right ascension and the declination of the target for the specified observing site, the distance between the target and the observing site, and the phase angle. Ephemeris files are downloaded in a csv format.   
\item \ul{Mutual orbit parameters.} The orbit of the satellite around the primary is a Keplerian mutual orbit defined by a set of parameters -- orbital period, semimajor axis, eccentricity, inclination, mean longitude, longitude of ascending node, and longitude of the periapsis -- at a certain epoch. This orbit is interpolated to calculate the position of the satellite at a certain time defined by the \texttt{Candela} user.    
\item \ul{Rotational and physical properties of the primary and secondary.} \texttt{Candela} can simulate a large variety of shapes; from the basic sphere, spheroid, and ellipsoid to the more complex figures of equilibrium for MacLaurin spheroid, Jacobi ellipsoid, and Roche binary \citep{Chandrasekhar1987}, and can also work with faceted shapes generated as .obj files. Once the user has selected the shapes for the primary and the secondary, the rotational and physical characteristics have to be specified for both of them. For each object, the user has to choose: size/axis length, density, rotational period, pole orientation, albedo, mass ratio for a Roche binary, and Hapke parameters (for example surface roughness, and amplitude of opposition surge among others). 
\end{enumerate}
The file paths (for ephemeris, output folder, and lightcurve file), the rotational and physical characteristics, as well as the mutual orbit parameters have to be specified in a csv generator to create a csv file that will be used as the input file for \texttt{Candela}. Using this input, the software will create a 3D rendering of the system and uses ray-tracing to generate a \textit{predicted lightcurve}. This predicted lightcurve is the expected brightness variation of the system as a function of time based on the system's characteristics reported in the csv file.

This predicted lightcurve can be compared to an \textit{observed lightcurve} obtained with telescopic facilities. Because some parameters have an associated error or are unknown, we employ a Markov Chain Monte Carlo (MCMC) procedure whose main goal is to derive the ideal set of parameters that best fits the observed lightcurve. In this case, we set a range of values with a starting point for \texttt{Candela} to find the best value within this range. The user is in charge of selecting the parameters that have to be fitted (all physical, rotational, and mutual orbit parameters can be fitted) as well as the number of MCMC iterations. As outputs, \texttt{Candela} will generate renders of the system, lightcurve plots, and files with Julian dates and expected magnitudes, as well as the best set(s) of parameters matching the predicted and observed lightcurves. We note that a complete MCMC analysis can be relevant if the inputs are sufficient and if all the parameters are tested. In our case, we emphasize that we only have one lightcurve obtained at one epoch for Logos, thus, our inputs are extremely limited. For example, we cannot probe the entire set of parameters, such as the spin axis orientation, as one lightcurve is insufficient to constrain this parameter. Similarly, our lack of knowledge regarding Zoe is restricting us to probe size and density for the components while matching the mutual orbit determination. Therefore, for the rest of this paper, we will present the best case scenario that matches our result, but it is important to keep in mind that we only have one ground-based dataset obtained at one epoch, and thus we present an illustrative result rather than a definitive result regarding this system. If mutual events are detected and/or if one can derive Zoe's lightcurve and/or a new lightcurve for Logos at a different epoch showing a change in the system's geometry, one will be able to propose a more definitive portrait of this system.

\begin{table*}[]
\caption{Characteristics of Logos assuming that it is a triaxial ellipsoid in hydrostatic equilibrium \citep{Chandrasekhar1987}.}
\label{tab:Jacobi}
\center
\begin{tabular}{|cllllll|}
\hline
\multicolumn{1}{|l|}{b/a} &
  \multicolumn{1}{l|}{c/a} &  
  \multicolumn{1}{l|}{Albedo} &  
  \multicolumn{1}{l|}{a {[}km{]}} &
  \multicolumn{1}{l|}{b {[}km{]}} &
  \multicolumn{1}{l|}{c {[}km{]}} &
  \multicolumn{1}{l|}{$\rho$ {[}g cm$^{-3}${]}}   \\ \hline
\multicolumn{7}{|c|}{$\xi$=90$^\circ$, H$^{LZ}_{MPC}$=6.87~mag}                                                                                                              \\ \hline
\multicolumn{1}{|c|}{\multirow{4}{*}{0.52}} &
  \multicolumn{1}{c|}{\multirow{4}{*}{0.40}} &
    \multicolumn{1}{l|}{0.04} &
  \multicolumn{1}{l|}{255} &
  \multicolumn{1}{l|}{133} &
  \multicolumn{1}{l|}{53} &
  \multicolumn{1}{l|}{$>$0.15}   \\ \cline{3-7} 
\multicolumn{1}{|c|}{} & \multicolumn{1}{c|}{} &     \multicolumn{1}{l|}{0.15} & \multicolumn{1}{l|}{131} & \multicolumn{1}{l|}{68} & \multicolumn{1}{l|}{27} & \multicolumn{1}{l|}{$>$0.15}   \\ \cline{3-7} 
\multicolumn{1}{|c|}{ } & \multicolumn{1}{c|}{ } &     \multicolumn{1}{l|}{0.20} & \multicolumn{1}{l|}{114}   & \multicolumn{1}{l|}{59} & \multicolumn{1}{l|}{24} & \multicolumn{1}{l|}{$>$0.15}   \\ \cline{3-7} 
\multicolumn{1}{|c|}{} & \multicolumn{1}{c|}{}   &  \multicolumn{1}{l|}{0.39$\pm$0.17} & \multicolumn{1}{l|}{82$^{+27}_{-14}$}   & \multicolumn{1}{l|}{42$^{+15}_{-7}$} & \multicolumn{1}{l|}{17$^{+6}_{-3}$} & \multicolumn{1}{l|}{$>$0.15}     \\ \hline
\multicolumn{7}{|c|}{$\xi$=60$^\circ$, H$^{LZ}_{MPC}$=6.87~mag}                                                                                                              \\ \hline
\multicolumn{1}{|l|}{\multirow{4}{*}{0.47}} &
  \multicolumn{1}{l|}{\multirow{4}{*}{0.37}} &
 \multicolumn{1}{l|}{0.04}  &  
  \multicolumn{1}{l|}{269} &
  \multicolumn{1}{l|}{127} &
  \multicolumn{1}{l|}{47} &
  \multicolumn{1}{l|}{$>$0.16}  \\ \cline{3-7} 
\multicolumn{1}{|l|}{} & \multicolumn{1}{l|}{}& \multicolumn{1}{l|}{0.15}  &   \multicolumn{1}{l|}{139}   & \multicolumn{1}{l|}{65} & \multicolumn{1}{l|}{24} & \multicolumn{1}{l|}{$>$0.16}  \\ \cline{3-7} 
\multicolumn{1}{|l|}{} & \multicolumn{1}{l|}{} &  \multicolumn{1}{l|}{0.20} & \multicolumn{1}{l|}{121}   & \multicolumn{1}{l|}{57} & \multicolumn{1}{l|}{21} & \multicolumn{1}{l|}{$>$0.16}    \\ \cline{3-7} 
\multicolumn{1}{|l|}{} & \multicolumn{1}{l|}{} &  \multicolumn{1}{l|}{0.39$\pm$0.17} & \multicolumn{1}{l|}{86$^{+29}_{-14}$}   & \multicolumn{1}{l|}{41$^{+13}_{-7}$} & \multicolumn{1}{l|}{15$^{+5}_{-2}$} & \multicolumn{1}{l|}{$>$0.16}   \\ \hline
\end{tabular}
\end{table*}

\subsection{Logos as a triaxial ellipsoid}
\label{sec:JacobiC}

As the reported ground-based lightcurve amplitude of Logos is not at or above the 0.9~mag threshold used to distinguish a contact binary from an elongated triaxial ellipsoid, as a first test case, we will consider that Logos is a triaxial ellipsoid in hydrostatic equilibrium. For a viewing angle of $\xi$=90$^\circ$ ($\xi$=60$^\circ$), the axis ratios of Logos are a/b=0.52 (0.47) and c/a=0.40 (0.37) with lower limit to the density of 0.15 (0.16)~g cm$^{-3}$ \citep{Chandrasekhar1987}. The characteristics of Logos as an ellipsoidal object and assuming several albedo estimates are summarized in Table~\ref{tab:Jacobi}. These characteristics, even approximate, give us a range of values and starting points for our modeling.  

We simulated a triaxial ellipsoid and we decided to fit for the a, b, and, c axes (with a$>$b$>$c, and object rotating along the c-axis), the density ($\rho$), and the albedo allowing \texttt{Candela} to retrieve the best set of parameters matching the observed lightcurve from Section~\ref{sec:LCtext}. The albedo range is between 0.04 and 0.60 which allows us to probe the entire range (and a bit more) of potential albedo values for this system with a starting point at 0.15 which is the typical albedo for a dynamically Cold Classical TNO and we selected a step size of 0.1. Even if we are not expecting a high density due to the object size and composition, the density ranges from 0.1 to 5~g cm$^{-3}$ with a starting point at 1~g cm$^{-3}$ and a step size of 0.1. The axes range from 50 to 300~km for a, 20 to 150~km for b, and 5 to 60~km for c with starting points at 130~km, 70~km, and 25~km for a, b, and c, respectively, and a step size of 1 for all three axes. One important parameter is the pole orientation of Logos, but, this parameter is unknown. For asteroids, it has been shown that several lightcurves at different epochs can be used to retrieve the pole orientation of an object \citep{Taylor1979, Hanus2011}. For simplicity and because we only have one lightcurve for Logos, we chose to not fit for pole orientation and we only considered objects viewed equator-on to avoid an over-interpretation of our limited dataset. A similar approach was used by \citet{Lacerda2014SQ317} for the modeling based on one lightcurve of the TNO (612620) 2003~SQ$_{317}$. We use the rotational period derived in Section~\ref{sec:LCtext}, and we do not fit for this parameter. 

  \begin{figure*}
  \includegraphics[width=18cm, angle=0]{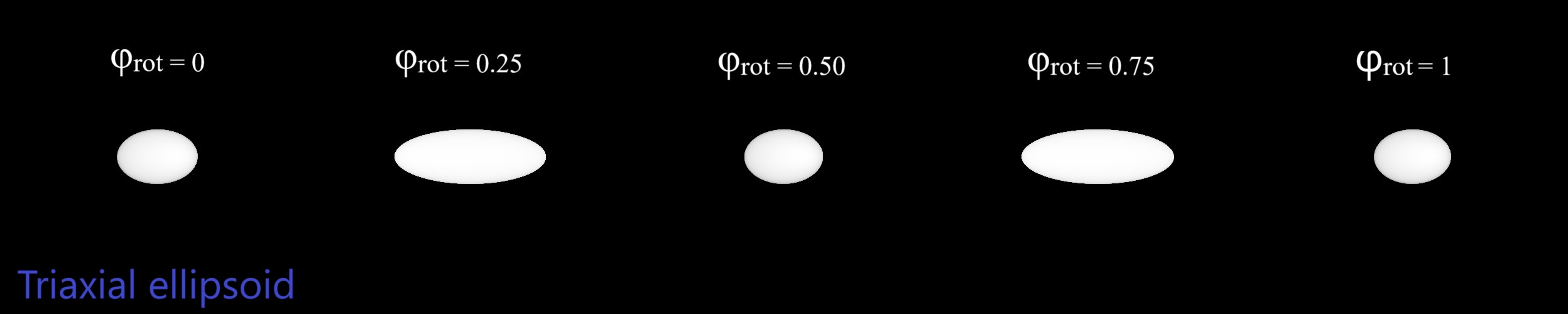} 
  \includegraphics[width=18cm, angle=0]{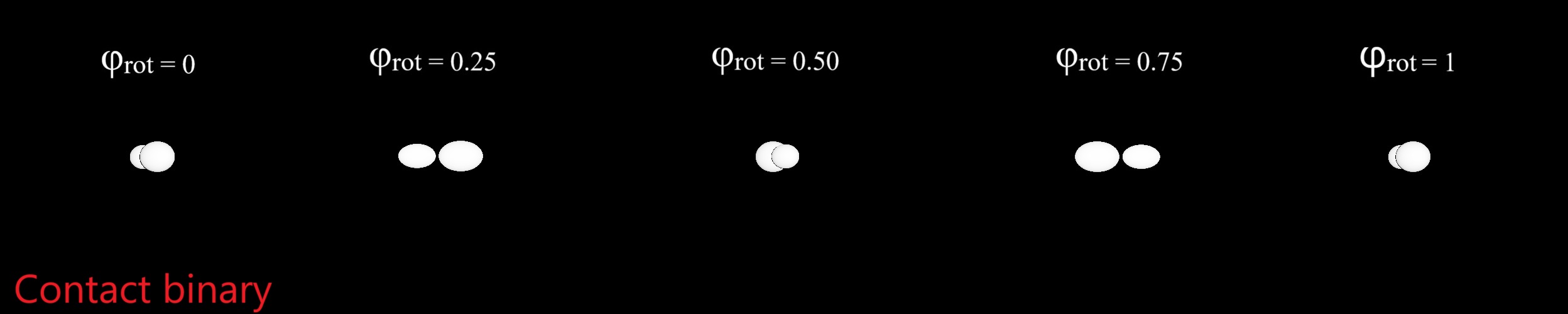}
  \includegraphics[width=10cm, angle=0]{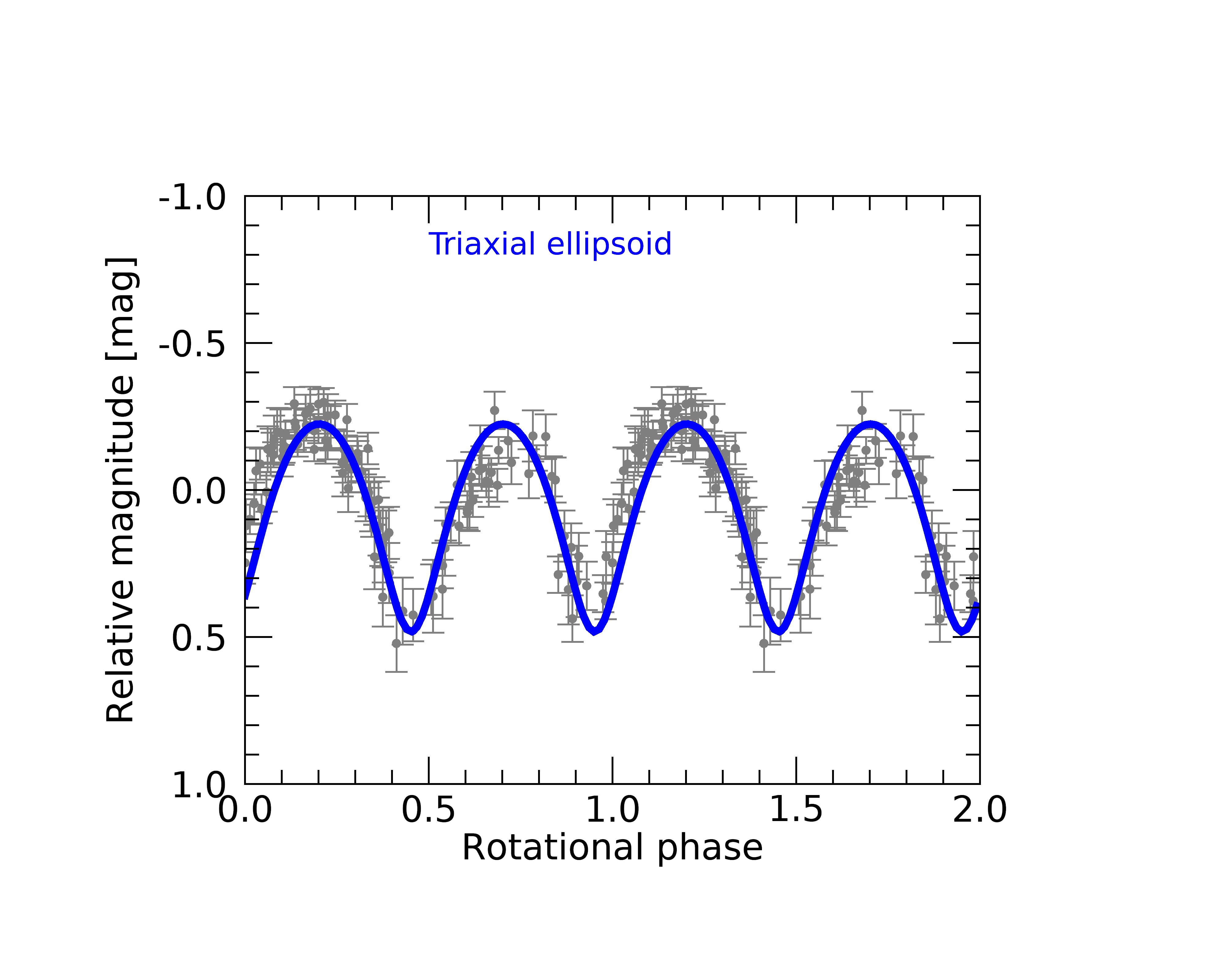}
  \includegraphics[width=10cm, angle=0]{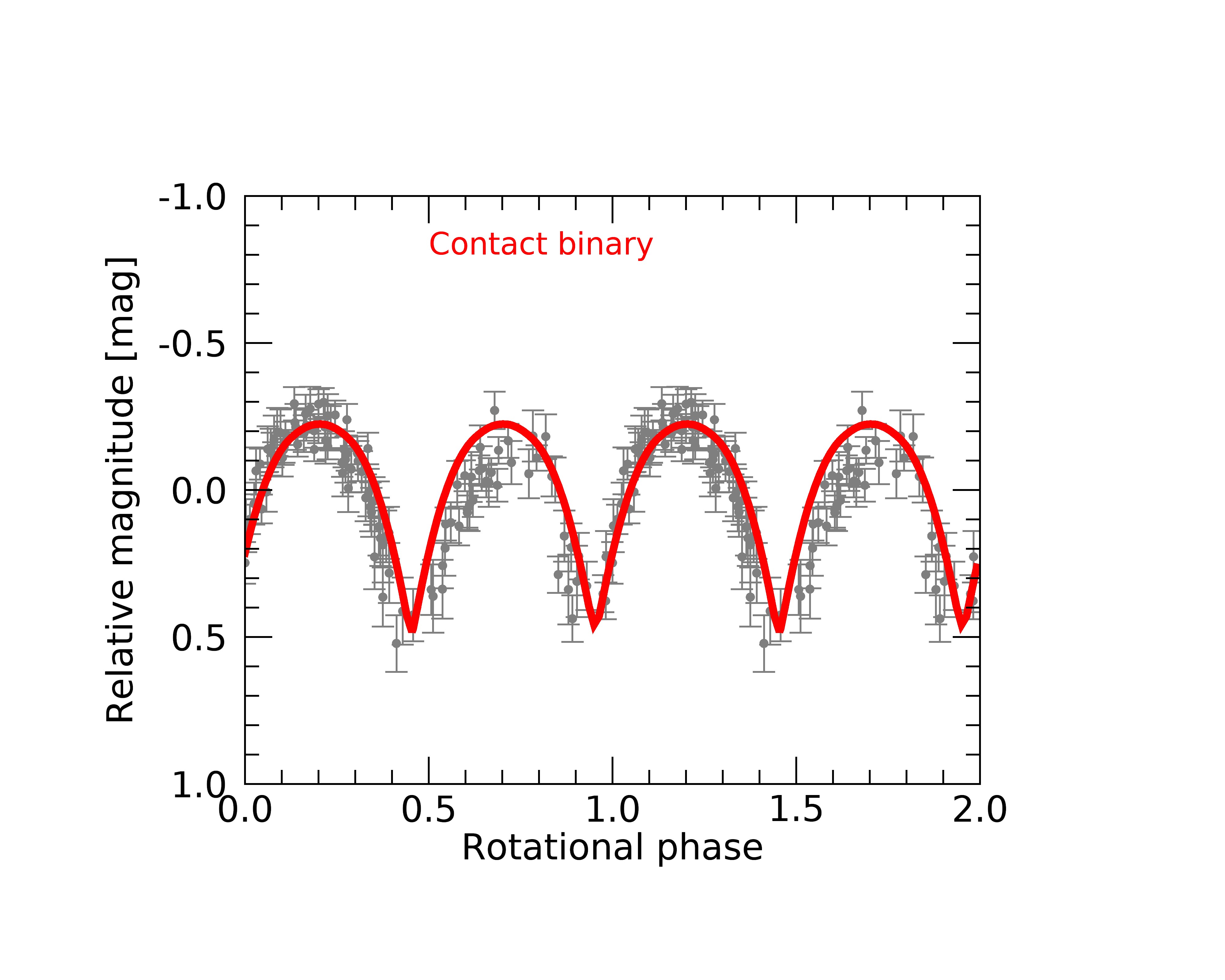}
  \caption{We present the triaxial ellipsoid and contact binary models that best fit the observed lightcurve. We selected a couple of screenshots to visualize the object's rotation at a rotational phase ($\upvarphi _{rot}$) of 0, 0.25, 0.50, 0.75, and 1 covering one full rotation. The overplotted continuous lines correspond to the best predicted lightcurves assuming a triaxial ellipsoid (blue) and a contact binary (red) fitting our observed lightcurve (grey circles corresponding to the observed data presented in Section~\ref{sec:LCtext}). Lightcurves are plotted over two full rotation (i.e., rotational phase, $\upvarphi _{rot}$ between 0 and 2.)  }
\label{fig:JacobiRocheCandela}
\end{figure*} 
 
\texttt{Candela} found that the best set of parameters fitting our observed lightcurve is:  a=159$\pm$23~km, b=70$\pm$10~km, c=25$\pm$15~km, $\rho$=1.0$\pm$0.5~g cm$^{-3}$, and albedo of 0.15$\pm$0.05. We emphasize that these parameters are highly correlated. For example, by adjusting the a and b axes of the triaxial, c will also be adjusted while still matching the ellipsoid equation. The fitted set of parameters suggests that if Logos is a triaxial ellipsoid, it is extremely elongated with a/b=2.27 which is close to the a/b=2.31 limit used for a triaxial object unstable due to rotational fission \citep{Jeans1919, Chandrasekhar1987}. The best predicted lightcurve based on this set of parameters is overplotted with our observed lightcurve in Figure~\ref{fig:JacobiRocheCandela}. The match between the observed and the predicted lightcurves is not perfect with a $\chi^2$=1.33. We can enumerate several reasons for this issue. We are unable to reproduce the asymmetry of the second maximum because in its current stage of development, \texttt{Candela} cannot handle albedo spot(s) nor irregular shapes. The predicted lightcurve is also unable to reproduce the wide inverted U-shape of the first maximum (as well as the second maximum to a certain degree but this maximum is also affected by the asymmetry), and this is an indication that we are not using the correct shape for our modeling.        

\subsection{Logos as a contact binary}
\label{sec:RocheC}

Since the observed lightcurve is not well interpreted by a triaxial ellipsoid, we next consider the case of Logos being a contact binary. We followed the same approach as in Section~\ref{sec:JacobiC} by fitting the axes, the density, and the albedo. For the same reason as previously discussed, only an equator-on configuration is considered. Inspired by the values reported in Table~\ref{tab:Leone}, we chose to probe for a$_{A}$ and b$_{A}$ from 30 to 50~km, c$_{A}$ between 20 and 50~km for Logos~A, and a$_{B}$, b$_{B}$, and c$_{B}$ between 20 and 50~km for the three axes of Logos~B. The ranges for the density and albedo as well as the step size for all parameters are the same as in Section~\ref{sec:JacobiC}. The mass ratio (q) is set to be between 0.1 and 1 with a starting point at 0.7. We use the rotational period derived in Section~\ref{sec:LCtext}, and we do not fit for this parameter.  

\texttt{Candela} inferred that a contact binary with a$_{A}$=40$\pm$15~km, b$_{A}$=30$\pm$12~km, c$_{A}$=20$\pm$12~km for Logos~A and a$_{B}$=40$\pm$15~km, b$_{B}$=34$\pm$12~km, c$_{B}$=33$\pm$12~km for Logos~B, as well as q=0.7$\pm$0.1, $\rho$=1.0$\pm$0.5~g/cm$^3$, and an albedo of 0.15$\pm$0.05. The contact binary corresponding to this optimal set of parameters and its associated predicted lightcurve are in Figure~\ref{fig:JacobiRocheCandela}. In this scenario, the 
$\chi^2$ of the observed and predicted lightcurve is 1.22. We point out that the asymmetry is still an issue, but the fit of the inverted U-shapes is greatly improved compared to the triaxial ellipsoid case. Based on lightcurve morphology (Section~\ref{sec:LCtext}) and modeling (Section~\ref{sec:interpretation}), we conclude that Logos is a close binary with a separation of 143$\pm$30~km between Logos~A and Logos~B.\\


\section{Logos-Zoe upcoming mutual events season}
\label{sec:mutualeventseason}
 
\citet{Grundy2011} inferred that the Logos-Zoe system should have a mutual event season centered on $\sim$2027. However, this prediction and modeling were made assuming that both Logos and Zoe have spherical shapes. Now that we have derived or at least constrained some rotational and physical properties of Logos and Zoe, we can update the mutual events prediction using the \texttt{Candela} software. 

For our prediction, we used the mutual orbit published by \citet{Grundy2011}. Still, we emphasize that its determination is based on \textit{HST} data obtained from 2001 to 2007 and assuming a Keplerian orbit. So, newly resolved astrometry would be useful to improve the mutual orbit determination as the uncertainty on the satellite position has grown significantly. Because the mutual orbit has not been updated recently, the timing precision of the mutual events is $\pm$1 week. We will consider several shapes (spheroid, ellipsoid, and contact binary) and periodicities for Zoe as our new observations have not allowed us to retrieve any information regarding its shape and period. As we do not expect Zoe to have significantly different physical properties than Logos (Logos and Zoe have similar surface colors, \citet{Benecchi2009}), we will assume that the density and albedo of each component are the same ($\rho_{Logos}$=$\rho_{Zoe}$=1~g/cm$^3$, and Albedo$_{Logos}$=Albedo$_{Zoe}$=0.15). Because the pole orientation of Zoe is also unknown, we will consider an equator-on configuration.

\subsection{Mutual events season}

As predicted by \citet{Grundy2011}, we confirm with the \texttt{Candela} software that the upcoming Logos-Zoe mutual events season will be between 2026 and 2029. Due to the long orbital period, only a handful of events happen per season. The complete list of the mutual events over the upcoming season can be found in Table~\ref{tab:events}. 

An \textit{inferior} mutual event happens when the satellite is in the background and a \textit{superior} mutual event when the primary is in the background. Some events depend highly on the objects' sizes. In some cases (see Table~\ref{tab:events}), events are not recorded with the by-default values but if Zoe is a bit larger than expected, an event may happen. Therefore, in Table~\ref{tab:events}, we listed the times when the objects are close to each other as well. 

Following, we will use the test case of the inferior event listed as 2027-3 in Table~\ref{tab:events} to illustrate the different shapes and rotational periods for Zoe. The modeling of the other events is in the Appendix. The same rotational and physical parameters have been used for all the events. For each mutual event, the brightness variation of the entire system is reported as a plot, but we emphasize that this brightness variation is the combination of the primary and secondary lightcurves as well as the mutual event. Disentangling the rotations of each component and the mutual event is required to isolate the mutual event.

\subsection{Dynamical considerations}
 
Before presenting the modeling of the upcoming mutual event season, we address some dynamical considerations regarding the Logos-Zoe system. As suggested in this paper, the Logos-Zoe system is a hierarchical system with three or even four objects. With such a highly complex system, it is possible that a large J2 and C22 can perturb the mutual orbit \citep{Proudfoot2024a, Proudfoot2024b, Nelsen2024}. Therefore, new astrometric observations are warranted to update the mutual orbit which can not be strictly Keplerian and thus could have precessed since the last \textit{HST} observations. Such a consideration will be discussed in greater detail in Proudfoot et al. (In preparation), therefore, we will not address it here.

\subsection{\underline{Option 1}: Zoe is a spheroid}
\label{sec:ZoeSphere}

As a first test, we consider Zoe a spherical object with a=b=c=30~km with an albedo of 0.15 and a density of 1~g/cm$^{3}$. Because Zoe is a sphere with a perfectly uniform surface (e.g., no albedo spot on the surface), its rotational lightcurve does not display any variability and therefore is perfectly flat. In this scenario, the rotational period of the object is irrelevant as the lightcurve is always flat.  

\begin{figure*}
\center
  \includegraphics[width=18cm, angle=0]{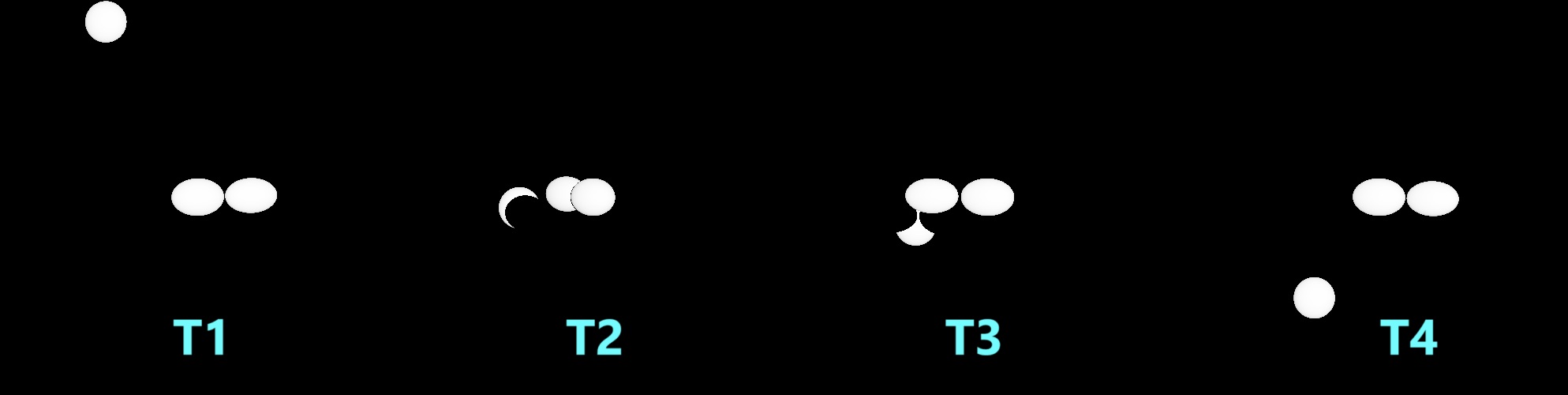}
  \includegraphics[width=10cm, angle=0]{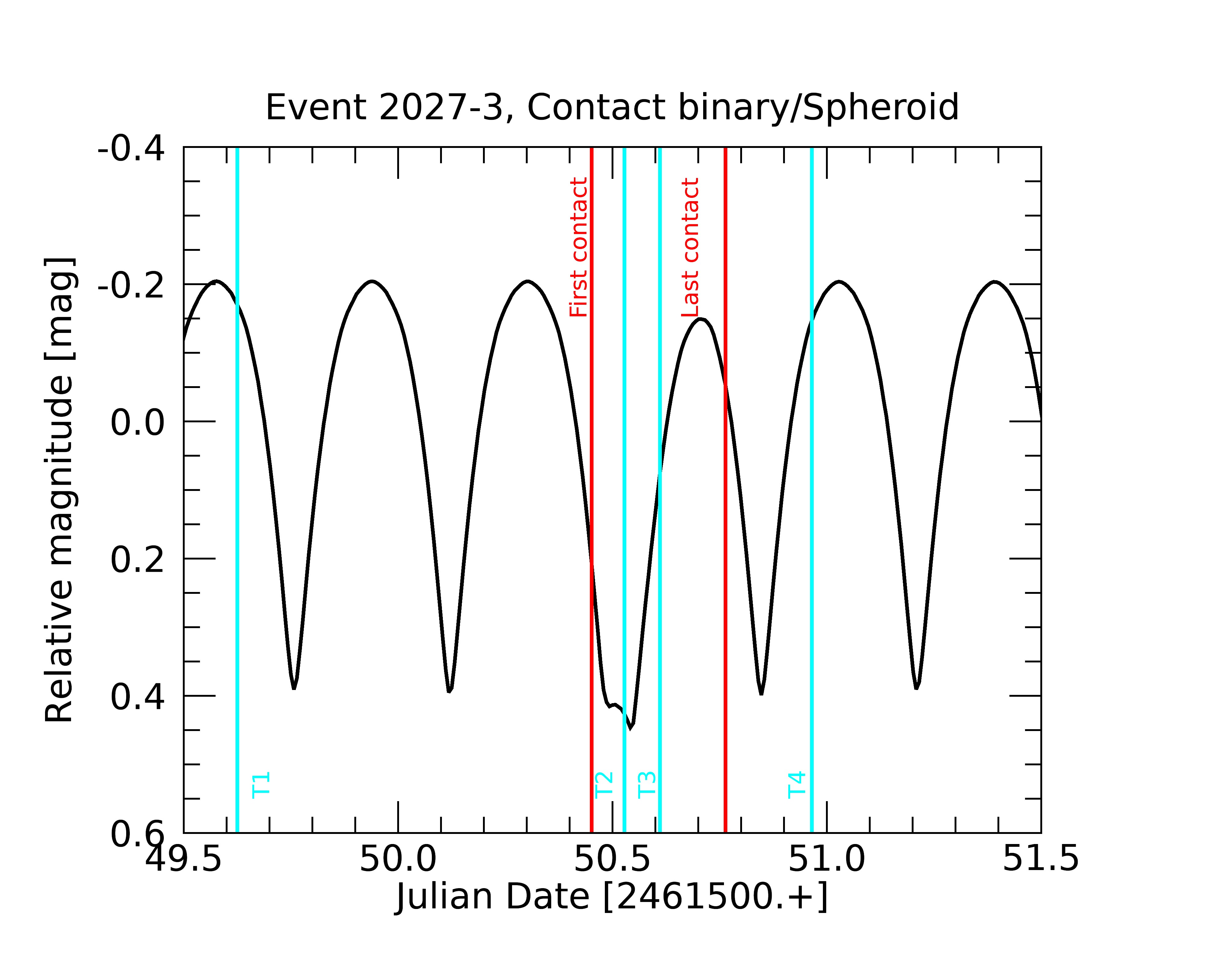}
  \caption{For Option 1, Zoe is a spherical object and Logos is a contact binary. In this Figure, we report four screenshots centered on Logos at different times (T1 to T4) to show the system configuration before (T1), during (T2 and T3), and after (T4) the 2027-3 mutual event (see Table~\ref{tab:events}). During this mutual event, Logos will cast its shadow on Zoe (inferior event). In the lower plot, we represent the brightness variation of the system as a function of time (black continuous line). The timings of the screenshots are indicated by the vertical cyan lines while the red vertical lines delimit the duration of the mutual event. The "first contact" red line indicates when the mutual event starts and the "last contact" red line indicates when the mutual event stops.   }
\label{fig:MutualEvent_Sphere}
\end{figure*}

In Figure~\ref{fig:MutualEvent_Sphere}, we propose four screenshots illustrating the system's geometry before (T1), during (T2 and T3), and after (T4) the mutual event when Logos' shadow is projected on Zoe. The lower part of Figure~\ref{fig:MutualEvent_Sphere} is the brightness variation of the system as a function of time with the superimposed mutual event. Because Zoe is a sphere, in this case, its rotational lightcurve is irrelevant, and we need to disentangle only the mutual event from the primary's lightcurve. The first and last contact corresponding to the start and end of the event are also plotted. Here, the event will last approximately 6~h.  

\subsection{\underline{Option 2}: Zoe is a triaxial ellipsoid}
\label{sec:ZoeJacobi}

\begin{figure*}
\center
  \includegraphics[width=18cm, angle=0]{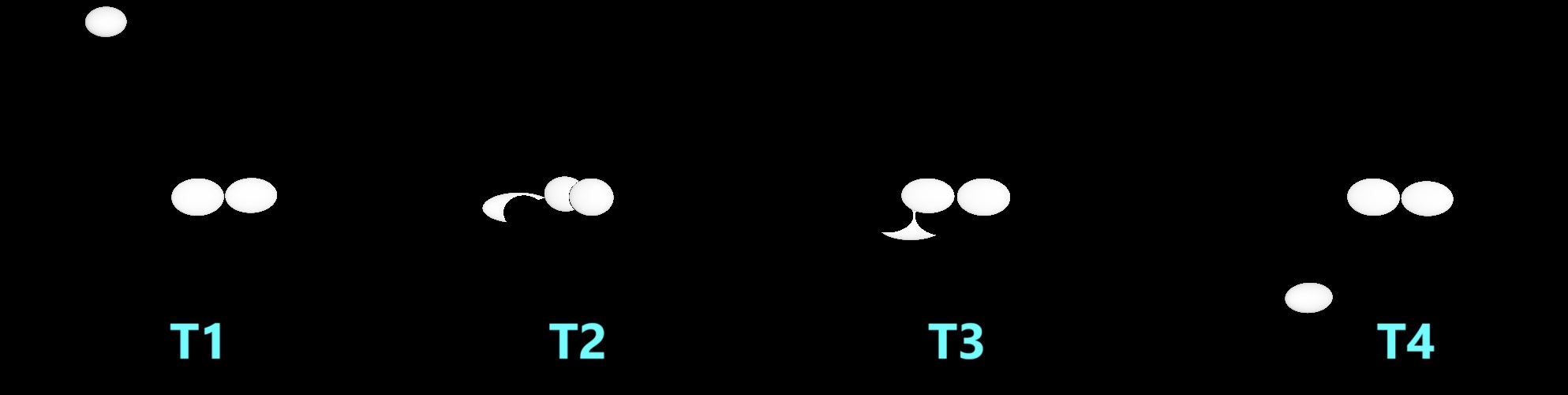}
  \includegraphics[width=10cm, angle=0]{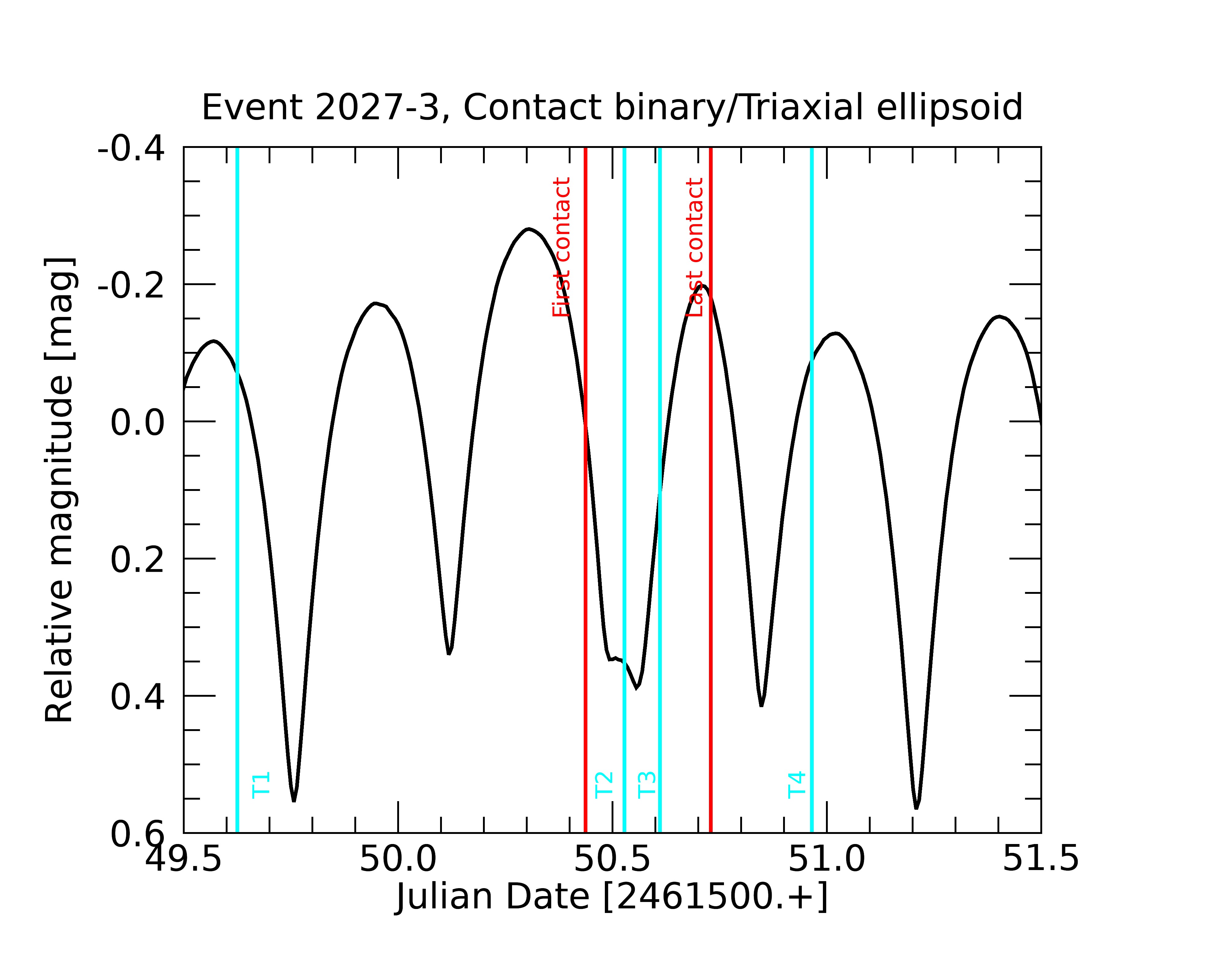}
  \caption{For Option 2, Zoe is a triaxial ellipsoid with a=52~km, b=30~km, and c=13~km and Logos is a contact binary. Zoe is rotating along its c-axis with a rotational period of 3~days. In this Figure, we report four screenshots centered on Logos at different times (T1 to T4) to show the system configuration before (T1), during (T2 and T3), and after (T4) the 2027-3 mutual event (see Table~\ref{tab:events}). During this mutual event, Logos will cast its shadow on Zoe (inferior event). In the lower plot, we represent the brightness variation of the system as a function of time (black continuous line). The timings of the screenshots are indicated by the vertical cyan lines while the red vertical lines delimit the duration of the mutual event. The "first contact" red line indicates when the mutual event starts and the "last contact" red line indicates when the mutual event stops.   }
\label{fig:MutualEvent_Jacobi}
\end{figure*}

Assuming that the variability of 0.6~mag noticed in the \textit{HST} datasets is the full lightcurve amplitude of Zoe, we estimate that the axis of Zoe are a=52~km, b=30~km and c=13~km with an albedo of 0.15. We select a density of 1~g/cm$^{3}$ and a by-default value of 3~days for Zoe's rotational period. This new case scenario is summarized in Figure~\ref{fig:MutualEvent_Jacobi}. 

As illustrated in the plot of Figure~\ref{fig:MutualEvent_Jacobi}, the mutual event needs to be disentangled from the lightcurve of the primary and the lightcurve of the secondary. The event will have approximately the same duration, but will start about 1~h earlier compared to the sphere case (Section~\ref{sec:ZoeSphere}).

\subsection{\underline{Option 3}: Zoe is a contact binary}

\begin{figure*}
\center
  \includegraphics[width=18cm, angle=0]{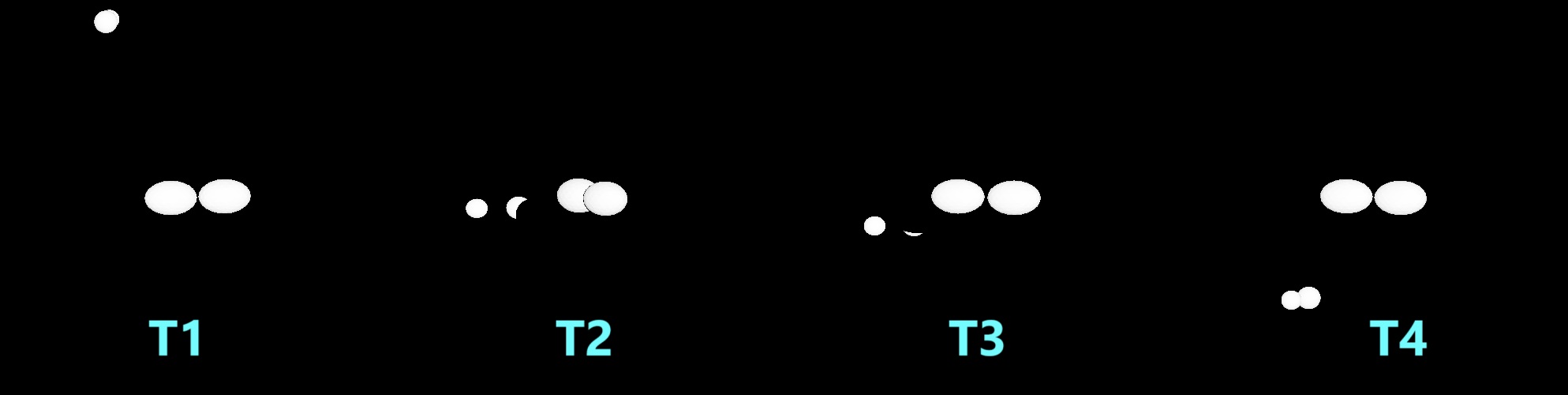}
  \includegraphics[width=10cm, angle=0]{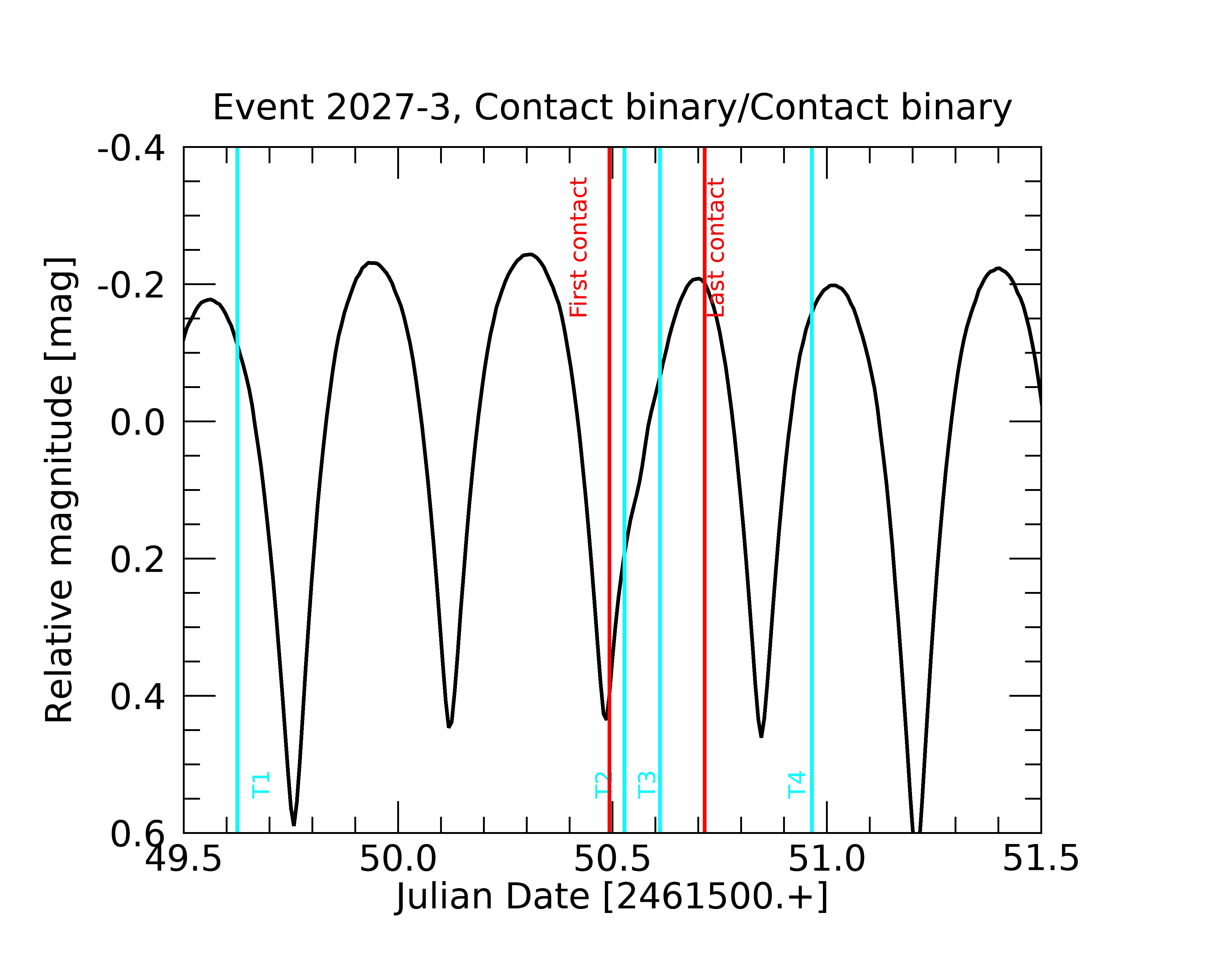}
  \caption{For Option 3, Zoe is a contact binary with semi-major axes of 18$\times$17$\times$16~km and 16$\times$15$\times$14~km and Logos is also a contact binary. Zoe is rotating with a rotational period of 3~days. In this Figure, we report four screenshots centered on Logos at different times (T1 to T4) to show the system configuration before (T1), during (T2 and T3), and after (T4) the 2027-3 mutual event (see Table~\ref{tab:events}). Logos will cast its shadow on Zoe (inferior event) during this mutual event. In the lower plot, we represent the brightness variation of the system as a function of time (black continuous line). The vertical cyan lines indicate the timings of the screenshots while the red vertical lines delimit the duration of the mutual event. The "first contact" red line indicates when the mutual event starts and the "last contact" red line indicates when the mutual event stops.   }
\label{fig:MutualEvent_Roche}
\end{figure*}

In this last case, we consider Zoe as a contact binary of two components whose sizes are 18$\times$17$\times$16~km and 16$\times$15$\times$14~km. We used again a density of 1~g/cm$^{3}$ and a rotational period of 3~days for Zoe. The contact binary case scenario is summarized in Figure~\ref{fig:MutualEvent_Roche}. In this case, the event will be less than 6~h compared to the other two cases. 

\subsection{Zoe's rotational periods}

Previously, we have demonstrated the variety of mutual events' morphology caused by Zoe's potential shapes. However, one has to consider the rotational period and so the rotational phase of Zoe which will also affect the modeling, the timing, and the morphology of the mutual events. 

\begin{figure*}
\center
  \includegraphics[width=15cm, angle=0]{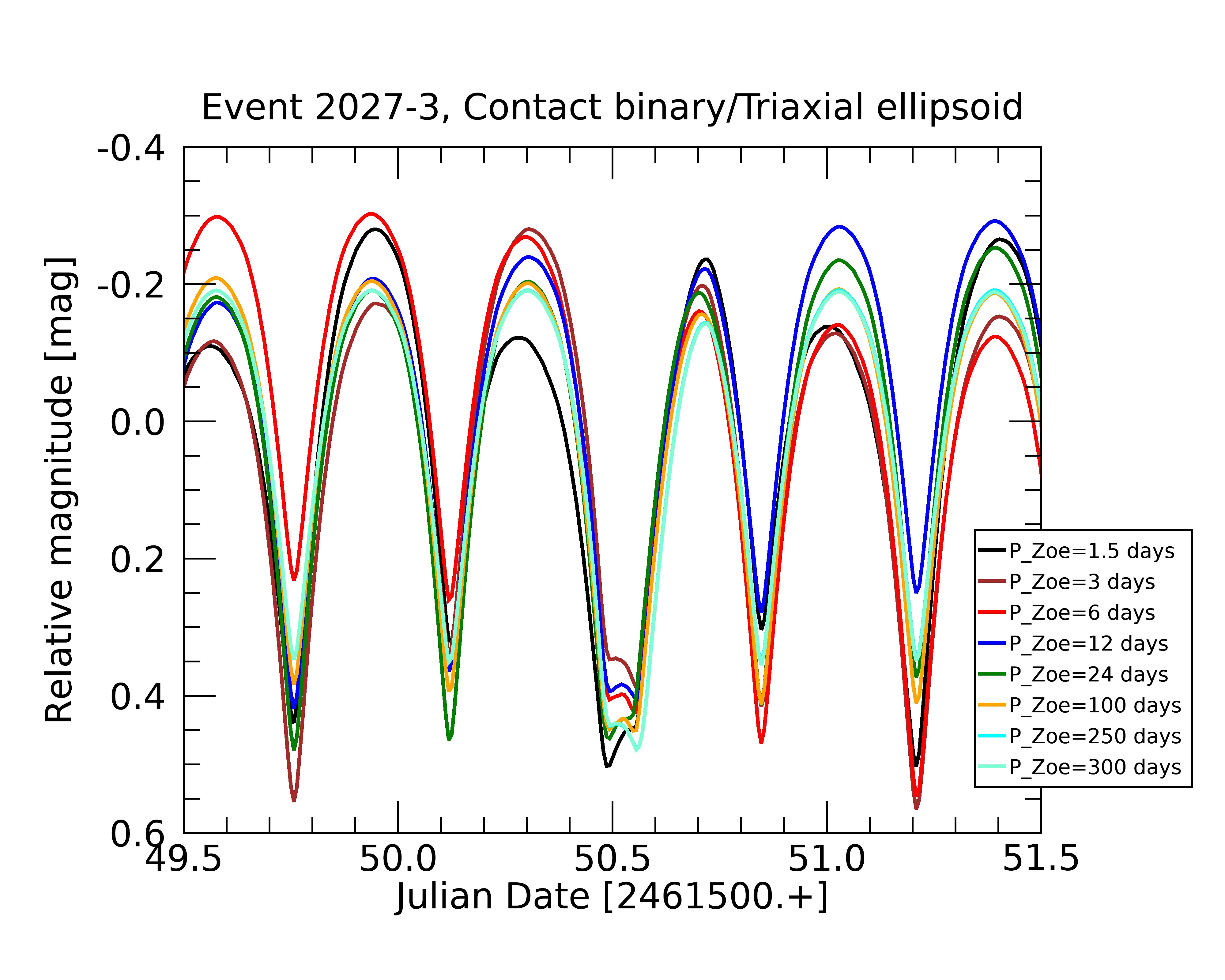}
  \caption{ We model the event 2027-3 assuming that Zoe is a triaxial ellipsoid with b/a=0.58 and c/a=0.42, but we use different rotational periods for Zoe. One can appreciate
that Zoe's rotational period dictates the morphology and depth of the event. }
\label{fig:MutualEvent_JacobiPeriod}
\end{figure*}

To illustrate the effect of Zoe's period, we consider the 2027-3 event assuming that Zoe is a triaxial ellipsoid with a range of rotational period from 1.5 to 300~days. For all instances, we use Zoe's rotational and physical characteristics described in Section~\ref{sec:ZoeJacobi}. This study is summarized in Figure~\ref{fig:MutualEvent_JacobiPeriod}. One can appreciate that the combined lightcurve (rotational lightcurves of the primary and the secondary and the mutual event) vary significantly with Zoe's rotational period. The combined lightcurves have different amplitudes and morphologies. Therefore, even if knowing the rotational period of Zoe in advance of the mutual event season to accurately predict the season, with the detection of at least one positive mutual event, one can constrain the rotational period of Zoe. We emphasize that even if only the triaxial ellipsoid case is reported here, a similar situation happens in the case of a contact binary.

\subsection{Notes on some events}

In Table~\ref{tab:events}, we report several close approaches. These approaches happen when the objects are very close to each other but no event is recorded. However, if the size of at least one of the objects is underestimated and/or if one of the objects is in a different rotational phase, an event may happen. The close approach listed as 2026-2 will happen on 2461345.49306 while the 2029-8 will be on 2462276.18056. 

The event 2028-6 is a double event (2028-6a and 2028-6b). In the first time, Zoe will be passing in front of Logos (2028-6a), and in the second time (2028-6b), Zoe will cast its shadow in one of Logos' components (see Appendix). We note that in the case of Contact binary/Contact binary, the 2028-6a event will be a close approach, but once again an event may happen if we have underestimated the size of the components. 

\begin{table*}[]
\caption{Estimated timings of Logos-Zoe mutual events are reported in this Table. We emphasize that due to the lack of updates since 2007 for the mutual orbit determination, the uncertainty on the mutual events timings is $\pm$1 week.}
\label{tab:events}
\resizebox{\textwidth}{!}{%
\begin{tabular}{|cccccc|}
\hline
\multicolumn{1}{|c|}{\multirow{2}{*}{Event ID}} &
  \multicolumn{1}{c|}{\multirow{2}{*}{\begin{tabular}[c]{@{}c@{}}Shape\\ Logos/Zoe\end{tabular}}} &
  \multicolumn{1}{c|}{\multirow{2}{*}{Start}} &
  \multicolumn{1}{c|}{\multirow{2}{*}{End}} &
  \multicolumn{1}{c|}{\multirow{2}{*}{Type of event}} &
  \multirow{2}{*}{Note} \\
\multicolumn{1}{|c|}{} &
  \multicolumn{1}{c|}{} &
  \multicolumn{1}{c|}{} &
  \multicolumn{1}{c|}{} &
  \multicolumn{1}{c|}{} &
   \\ \hline
\multicolumn{6}{|c|}{\underline{Year}: 2026} \\ \hline
\multicolumn{1}{|c|}{\multirow{3}{*}{2026-1}} &
  \multicolumn{1}{c|}{Contact binary/Spheroid} &
  \multicolumn{1}{c|}{2461240.47917} & 
  \multicolumn{1}{c|}{2461240.66667} & 
  \multicolumn{1}{c|}{Inferior} &  
   \\ \cline{2-6} 
\multicolumn{1}{|c|}{} &
  \multicolumn{1}{c|}{Contact binary/Triaxial ellipsoid} &
  \multicolumn{1}{c|}{ 2461240.48611} & 
  \multicolumn{1}{c|}{ 2461240.66667   } & 
  \multicolumn{1}{c|}{Inferior} & %
   \\ \cline{2-6} 
\multicolumn{1}{|c|}{} &
  \multicolumn{1}{c|}{Contact binary/Contact binary} &
  \multicolumn{1}{c|}{no event} & 
  \multicolumn{1}{c|}{no event} & 
  \multicolumn{1}{c|}{} &
   Close approach  \\ \hline
\multicolumn{1}{|c|}{\multirow{3}{*}{2026-2}} &
  \multicolumn{1}{c|}{Contact binary/Spheroid} &
  \multicolumn{1}{c|}{no event} & 
  \multicolumn{1}{c|}{no event} & 
  \multicolumn{1}{c|}{ } &
   Close approach  
   \\ \cline{2-6} 
\multicolumn{1}{|c|}{} &
  \multicolumn{1}{c|}{Contact binary/Triaxial ellipsoid} &  
 \multicolumn{1}{c|}{no event} & 
  \multicolumn{1}{c|}{no event} & 
  \multicolumn{1}{c|}{ } &
   Close approach  
   \\ \cline{2-6} 
\multicolumn{1}{|c|}{} &
  \multicolumn{1}{c|}{Contact binary/Contact binary} &
  \multicolumn{1}{c|}{no event} & 
  \multicolumn{1}{c|}{no event} & 
  \multicolumn{1}{c|}{ } &
   Close approach  
 \\ \hline
\multicolumn{6}{|c|}{ \underline{Year}: 2027} \\ \hline
\multicolumn{1}{|c|}{\multirow{3}{*}{2027-3}} &
  \multicolumn{1}{c|}{Contact binary/Spheroid} &
  \multicolumn{1}{c|}{2461550.45139} & 
  \multicolumn{1}{c|}{2461550.76389} & 
  \multicolumn{1}{c|}{Inferior} &
   \\ \cline{2-6} 
\multicolumn{1}{|c|}{} &
  \multicolumn{1}{c|}{Contact binary/Triaxial ellipsoid} &
  \multicolumn{1}{c|}{2461550.43750} &  
  \multicolumn{1}{c|}{2461550.72917 } &  
  \multicolumn{1}{c|}{ Inferior} &
   \\ \cline{2-6} 
\multicolumn{1}{|c|}{} &
  \multicolumn{1}{c|}{Contact binary/Contact binary} &
  \multicolumn{1}{c|}{2461550.49306} & 
  \multicolumn{1}{c|}{ 2461550.71528} & 
  \multicolumn{1}{c|}{Inferior} &
 \\ \hline
\multicolumn{1}{|c|}{\multirow{3}{*}{2027-4}} &
  \multicolumn{1}{c|}{Contact binary/Spheroid} &
  \multicolumn{1}{c|}{2461655.43750 } &
  \multicolumn{1}{c|}{2461655.79167} &
  \multicolumn{1}{c|}{ Superior} &
   \\ \cline{2-6} 
\multicolumn{1}{|c|}{} &
  \multicolumn{1}{c|}{Contact binary/Triaxial ellipsoid} &
  \multicolumn{1}{c|}{2461655.28472} &
  \multicolumn{1}{c|}{2461655.88889} &
  \multicolumn{1}{c|}{Superior } &
   \\ \cline{2-6} 
\multicolumn{1}{|c|}{} &
  \multicolumn{1}{c|}{Contact binary/Contact binary} &
  \multicolumn{1}{c|}{2461655.25000 } &
  \multicolumn{1}{c|}{2461656.06250} &
  \multicolumn{1}{c|}{Superior } &
   \\ \hline
\multicolumn{6}{|c|}{\underline{Year}: 2028} \\ \hline
\multicolumn{1}{|c|}{\multirow{3}{*}{2028-5}} &
  \multicolumn{1}{c|}{Contact binary/Spheroid} &
  \multicolumn{1}{c|}{2461860.34028} &
  \multicolumn{1}{c|}{2461860.70139 } &%
  \multicolumn{1}{c|}{Inferior} &
   \\ \cline{2-6} 
\multicolumn{1}{|c|}{} &
  \multicolumn{1}{c|}{Contact binary/Triaxial ellipsoid} &
  \multicolumn{1}{c|}{2461860.36806} &
  \multicolumn{1}{c|}{ 2461860.68750} &
  \multicolumn{1}{c|}{ Inferior} &
   \\ \cline{2-6} 
\multicolumn{1}{|c|}{} &
  \multicolumn{1}{c|}{Contact binary/Contact binary} &
  \multicolumn{1}{c|}{2461860.38194 } &
  \multicolumn{1}{c|}{2461860.67361 } &
  \multicolumn{1}{c|}{Inferior} &
 \\ \hline
\multicolumn{1}{|c|}{\multirow{3}{*}{2028-6a}} &
  \multicolumn{1}{c|}{Contact binary/Spheroid} &
  \multicolumn{1}{c|}{2461964.90278} &%
  \multicolumn{1}{c|}{2461965.09028} &%
  \multicolumn{1}{c|}{ Superior} &  Double event
   \\ \cline{2-6} 
\multicolumn{1}{|c|}{} &
  \multicolumn{1}{c|}{Contact binary/Triaxial ellipsoid} &
  \multicolumn{1}{c|}{2461964.89583} &%
  \multicolumn{1}{c|}{2461965.09028} &%
  \multicolumn{1}{c|}{Superior} &   Double event
   \\ \cline{2-6} 
\multicolumn{1}{|c|}{} &
  \multicolumn{1}{c|}{Contact binary/Contact binary} &
  \multicolumn{1}{c|}{ no event} &
  \multicolumn{1}{c|}{ no event} &
  \multicolumn{1}{c|}{ } & Close approach
   \\ \hline
   \multicolumn{1}{|c|}{\multirow{3}{*}{2028-6b}} &
  \multicolumn{1}{c|}{Contact binary/Spheroid} &
  \multicolumn{1}{c|}{2461966.02083} &%
  \multicolumn{1}{c|}{2461966.15278} &%
  \multicolumn{1}{c|}{ Superior} &  Double event
   \\ \cline{2-6} 
\multicolumn{1}{|c|}{} &
  \multicolumn{1}{c|}{Contact binary/Triaxial ellipsoid} &
  \multicolumn{1}{c|}{2461964.00694} &%
  \multicolumn{1}{c|}{2461966.18750} &%
  \multicolumn{1}{c|}{Superior} &   Double event
   \\ \cline{2-6} 
\multicolumn{1}{|c|}{} &
  \multicolumn{1}{c|}{Contact binary/Contact binary} &
  \multicolumn{1}{c|}{ 2461966.00694} &
  \multicolumn{1}{c|}{ 2461966.17361} &
  \multicolumn{1}{c|}{Superior } &  
   \\ \hline
\multicolumn{6}{|c|}{\underline{Year}: 2029} \\ \hline
\multicolumn{1}{|c|}{\multirow{3}{*}{2029-7}} &
  \multicolumn{1}{c|}{Contact binary/Spheroid} &
  \multicolumn{1}{c|}{2462170.38889} &
  \multicolumn{1}{c|}{2462170.61806} &
  \multicolumn{1}{c|}{Inferior} &
   \\ \cline{2-6} 
\multicolumn{1}{|c|}{} &
  \multicolumn{1}{c|}{Contact binary/Triaxial ellipsoid} &
  \multicolumn{1}{c|}{2462170.38889} &
  \multicolumn{1}{c|}{2462170.61111} &
  \multicolumn{1}{c|}{ Inferior} &
   \\ \cline{2-6} 
\multicolumn{1}{|c|}{} &
  \multicolumn{1}{c|}{Contact binary/Contact binary} &
  \multicolumn{1}{c|}{2462170.45139} &
  \multicolumn{1}{c|}{2462170.54861} &
  \multicolumn{1}{c|}{Inferior} &
 \\ \hline
\multicolumn{1}{|c|}{\multirow{3}{*}{2029-8}} &
  \multicolumn{1}{c|}{Contact binary/Spheroid} &
  \multicolumn{1}{c|}{ no event} &
  \multicolumn{1}{c|}{ no event} &
  \multicolumn{1}{c|}{ } & Close approach
   \\ \cline{2-6} 
\multicolumn{1}{|c|}{} &
  \multicolumn{1}{c|}{Contact binary/Triaxial ellipsoid} &
  \multicolumn{1}{c|}{ no event} &
  \multicolumn{1}{c|}{ no event} &
  \multicolumn{1}{c|}{ } &Close approach
   \\ \cline{2-6} 
\multicolumn{1}{|c|}{} &
  \multicolumn{1}{c|}{Contact binary/Contact binary} &
  \multicolumn{1}{c|}{ no event} &
  \multicolumn{1}{c|}{ no event} &
  \multicolumn{1}{c|}{ } &Close approach
   \\ \hline
\end{tabular}%
}
\end{table*}

\section{Conclusion} 
\label{sec:conclusion}

Based on space- and ground-based observations obtained over more than two decades, we conclude the following regarding the Logos-Zoe system:

\begin{itemize}
\item The primary, Logos, rotates in 17.43$\pm$0.06~h with a lightcurve amplitude of 0.70$\pm$0.07~mag. We infer that Logos is a likely close/contact binary based on the lightcurve morphology and large variability. Using the \texttt{Candela} software, we model Logos as a close binary with a mass ratio q=0.7 and a density $\rho$=1~g/cm$^3$. 
\item The secondary, Zoe, is probably a (very) slow rotator with a rotational period from days to months. Zoe's shape is still an open question. 
\item We infer that the Logos-Zoe system is asynchronous with P$_{orb}$$\neq$P$_{Logos}$$\neq$P$_{Zoe}$.       
\item The Logos-Zoe system undergoes mutual events with its components passing in front or behind each other every $\sim$151 years. The upcoming mutual events season will start in 2026 and end in 2029. Due to the long orbital period of $\sim$310 days, there will be only up to two events per year.
\item Using the \texttt{Candela} software, we modeled the entire upcoming mutual events season and its events. Because the rotational period and shape of Zoe are unknown, we considered three shapes (spherical, elongated, and contact binary), and we illustrated the effects of the rotational period (and so rotational phase) of Zoe on the events morphology. 
\item With the current mutual orbit determination. the timing uncertainty on the individual events is $\pm$1 week. Therefore, new astrometry to update and improve the mutual orbit determination is highly desired. 
\item Our modeling is based on our current knowledge of the system and can be updated as our understanding of the system evolves. We also note that the pole orientation of Logos and Zoe will affect the modeling and prediction. Unfortunately, because we have no estimate nor constraints for their pole orientations, we did not explore this parameter space. We only consider equatorial viewing cases to avoid an overinterpretation of the ground- and space-based datasets. Potential precession effects have not been taken into account.   
\item Positive detections of inferior and superior events over the mutual events season will be crucial to improving the modeling of the system and retrieving the rotational and physical properties of Logos and Zoe.
\item Mutual event prediction and detection is an efficient technique to derive physical properties (size, shape, density, and, albedo of each component) for binaries/multiples but it is still fairly underutilized to study the trans-Neptunian belt.
\item Finally, we point out that retrieving basic pieces of information about the rotational and physical properties of binary/multiple systems, but also single trans-neptunian objects, provides clues and constraints on the early stages of our Solar System. 
\end{itemize}


\acknowledgments

This paper includes data gathered with the 6.5~m Magellan-Baade Telescope located at Las Campanas Observatory, Chile. This research is based on data obtained at the Lowell Discovery Telescope (LDT). Lowell Observatory is a private, non-profit institution dedicated to astrophysical research and public appreciation of astronomy and operates the LDT in partnership with Boston University, the University of Maryland, the University of Toledo, Northern Arizona University, and Yale University. Partial support of the LDT was provided by Discovery Communications. LMI was built by Lowell Observatory using funds from the National Science Foundation (AST-1005313). We are grateful to the Magellan and LDT staffs.  \\
The authors acknowledge the contribution of several Capstone students from Northern Arizona University’s (NAU) School of Informatics, Computing, and Cyber Systems to the development of the \texttt{licht} software; in Year 1: Zachary Kramer, Brian Donnelly, Matt Rittenback, in Year 2: Bradley Kukuk, Jessica Smith, John Jacobelli, Matthew Amato-Yarbrough, Batai Finley, in Year 3: Andres Sepulveda, Brandon Visoky, Eleanor Carlos, and Reyna Orendain. The \texttt{Candela} software has been developed by Felicity Escarzaga with support from Brian Donnelly.\\
Authors made use of the JPL Horizons Systems which was developed at the Jet Propulsion Laboratory (Solar System Dynamics Group), California Institute of Technology, under contract with the National Aeronautics and Space Administration.\\
AT and SSS acknowledge support from the National Science Foundation with grant 1734484 awarded to the ``Comprehensive Study of the Most Pristine Objects Known in the Outer Solar System'' and grant 2109207 awarded to the ``Resonant Contact Binaries in the Trans-Neptunian Belt''. AT, KSN, SSS, WMG, and FE acknowledge support from NASA-Solar System Observations (SSO) with grant 80NSSC22K0659 awarded to ``Mutual events in the trans-Neptunian belt". \\

%

\vspace{5mm}
\facilities{LDT, Magellan:Baade}





\clearpage
\newpage
\bibliography{biblio}{}
\bibliographystyle{aasjournal}



\newpage
  \appendix
\label{sec:Appendix}

The Logos-Zoe mutual events in the upcoming season are reported in this Appendix (2026-1 to 2029-7, except 2027-3 which is discussed in the paper). As for the event labeled 2027-3, we consider several shapes for Zoe. 


  \begin{figure*}[ht!]
\center
  \includegraphics[width=18cm, angle=0]{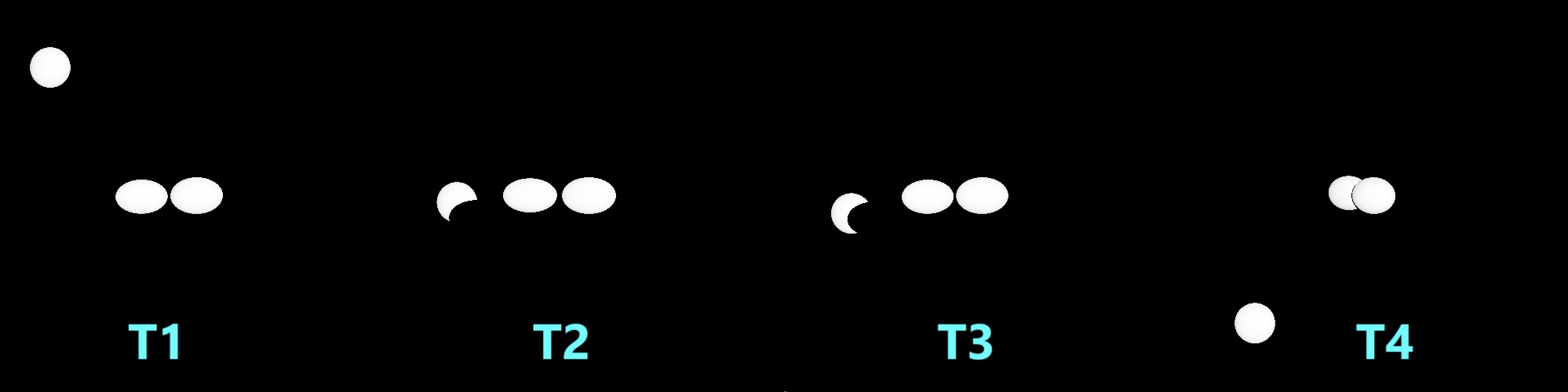}
   \includegraphics[width=10cm, angle=0]{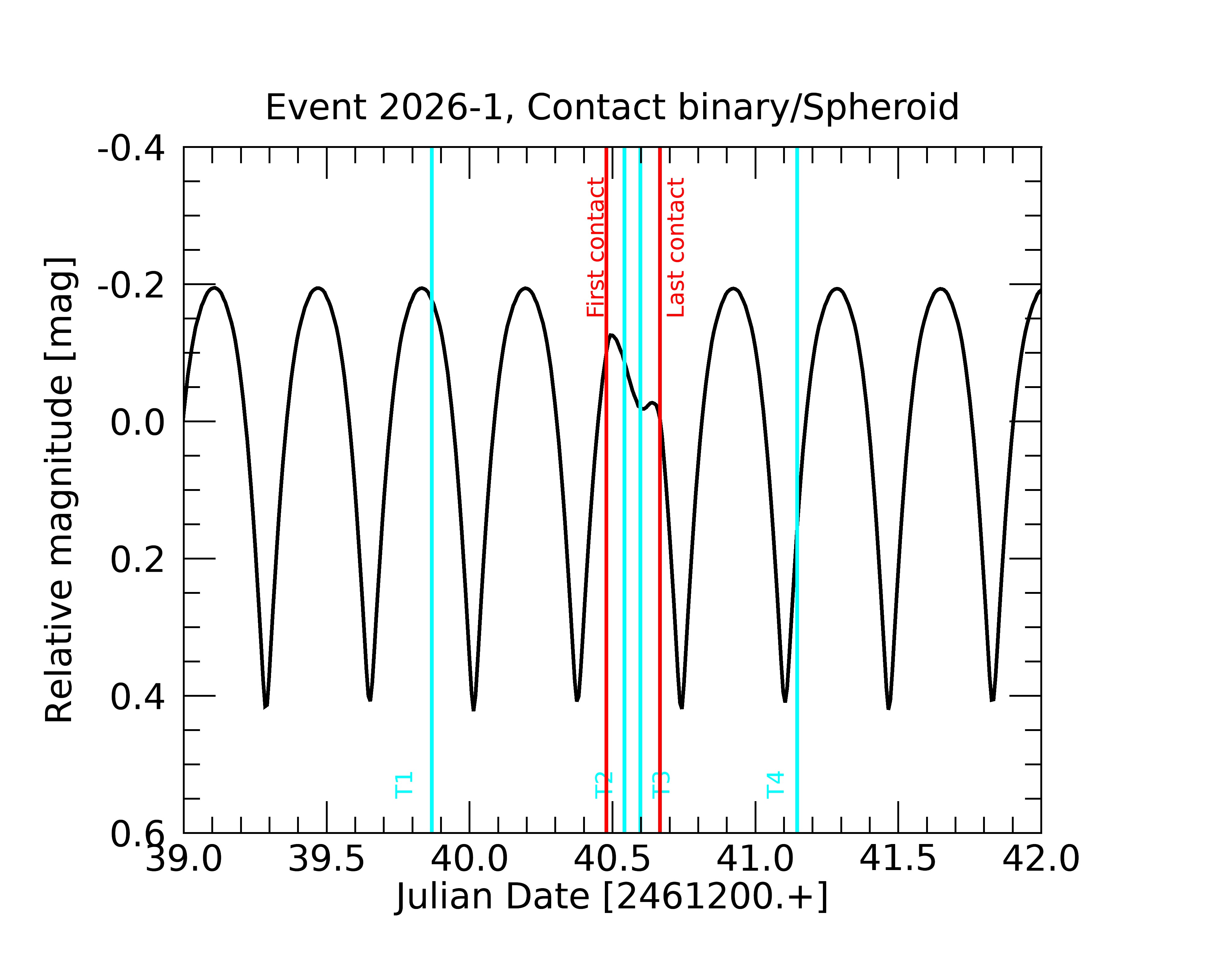}
  \caption{Modeling of the 2026-1 event assuming that Zoe is a spherical object.}
\label{fig:event2026-1-sphere}
\end{figure*} 


  \begin{figure*}[ht!]
\center
  \includegraphics[width=18cm, angle=0]{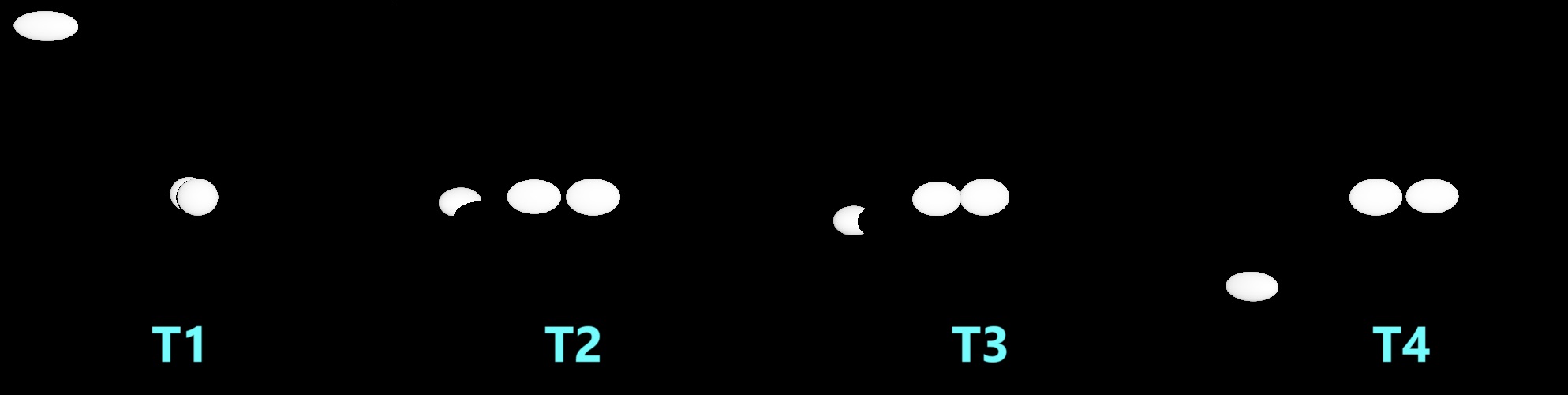}
   \includegraphics[width=10cm, angle=0]{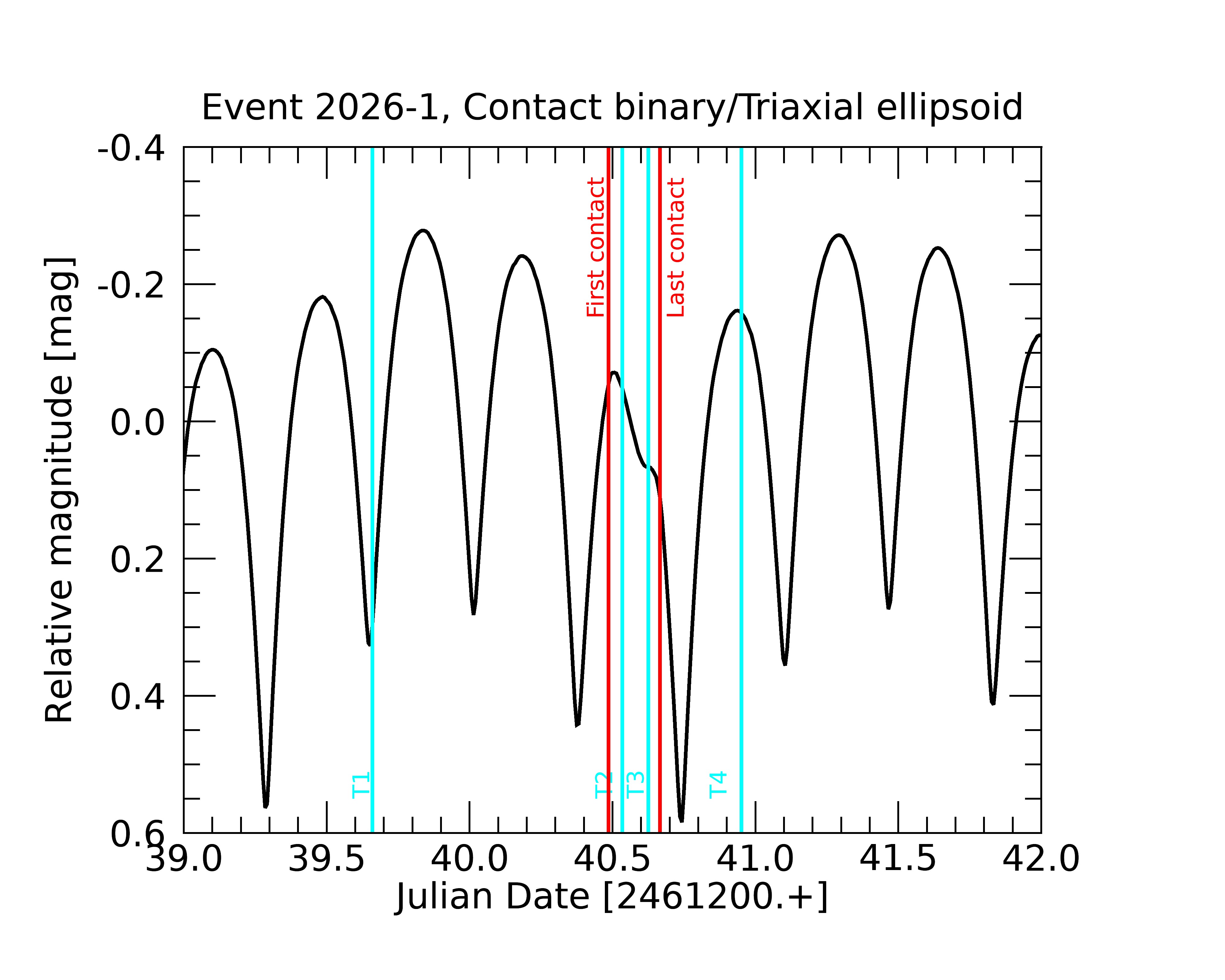}
  \caption{Modeling of the 2026-1 event assuming that Zoe is an ellipsoidal object.}
\label{fig:event2026-1-jacobi}
\end{figure*} 


 


  \begin{figure*}[ht!]
\center
  \includegraphics[width=18cm, angle=0]{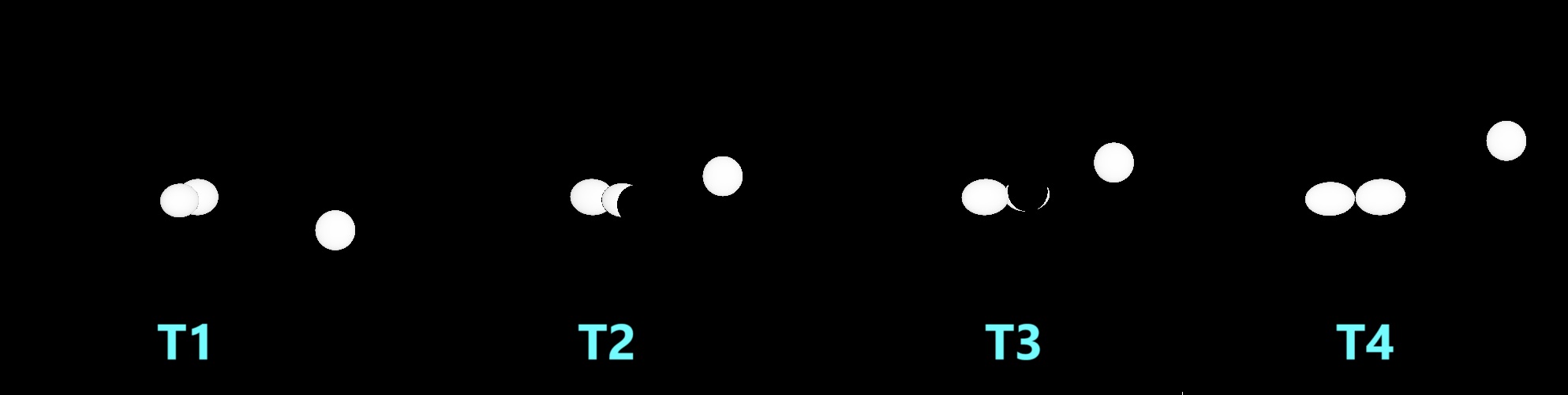}
   \includegraphics[width=10cm, angle=0]{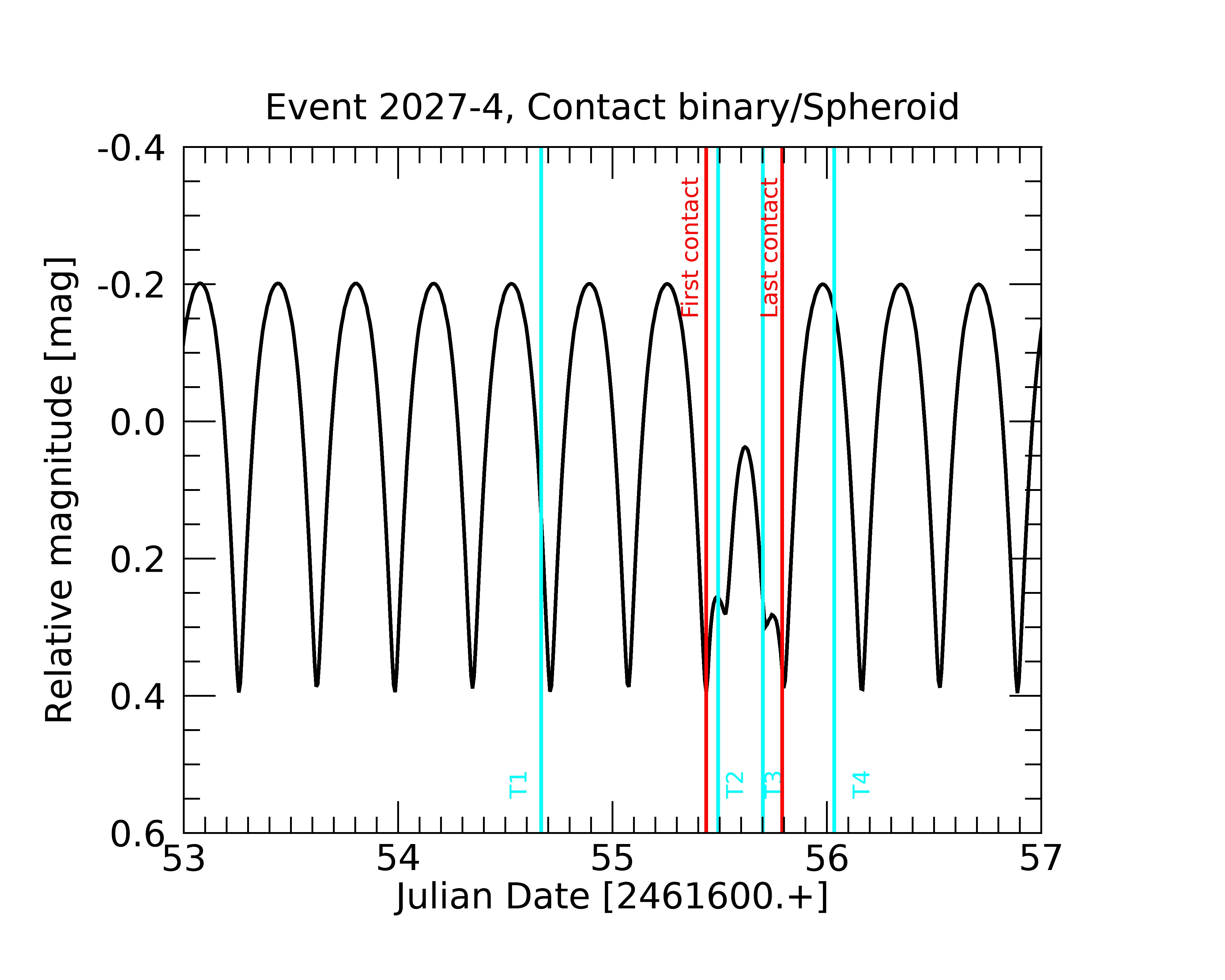}
  \caption{Modeling of the 2027-4 event assuming that Zoe is a spherical object.}
\label{fig:event2027-4-sphere}
\end{figure*} 


  \begin{figure*}[ht!]
\center
  \includegraphics[width=18cm, angle=0]{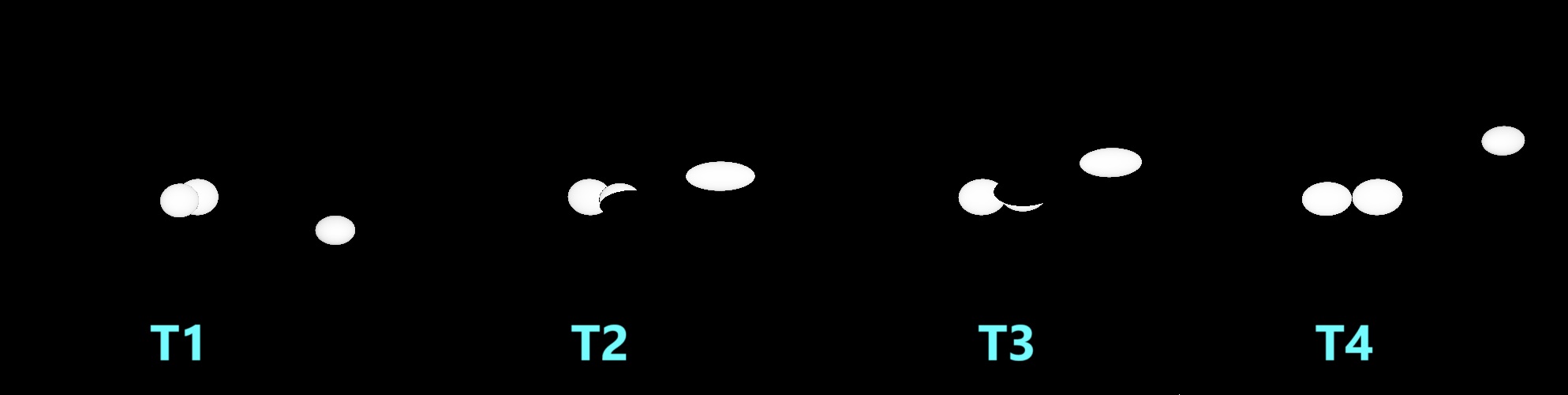}
   \includegraphics[width=10cm, angle=0]{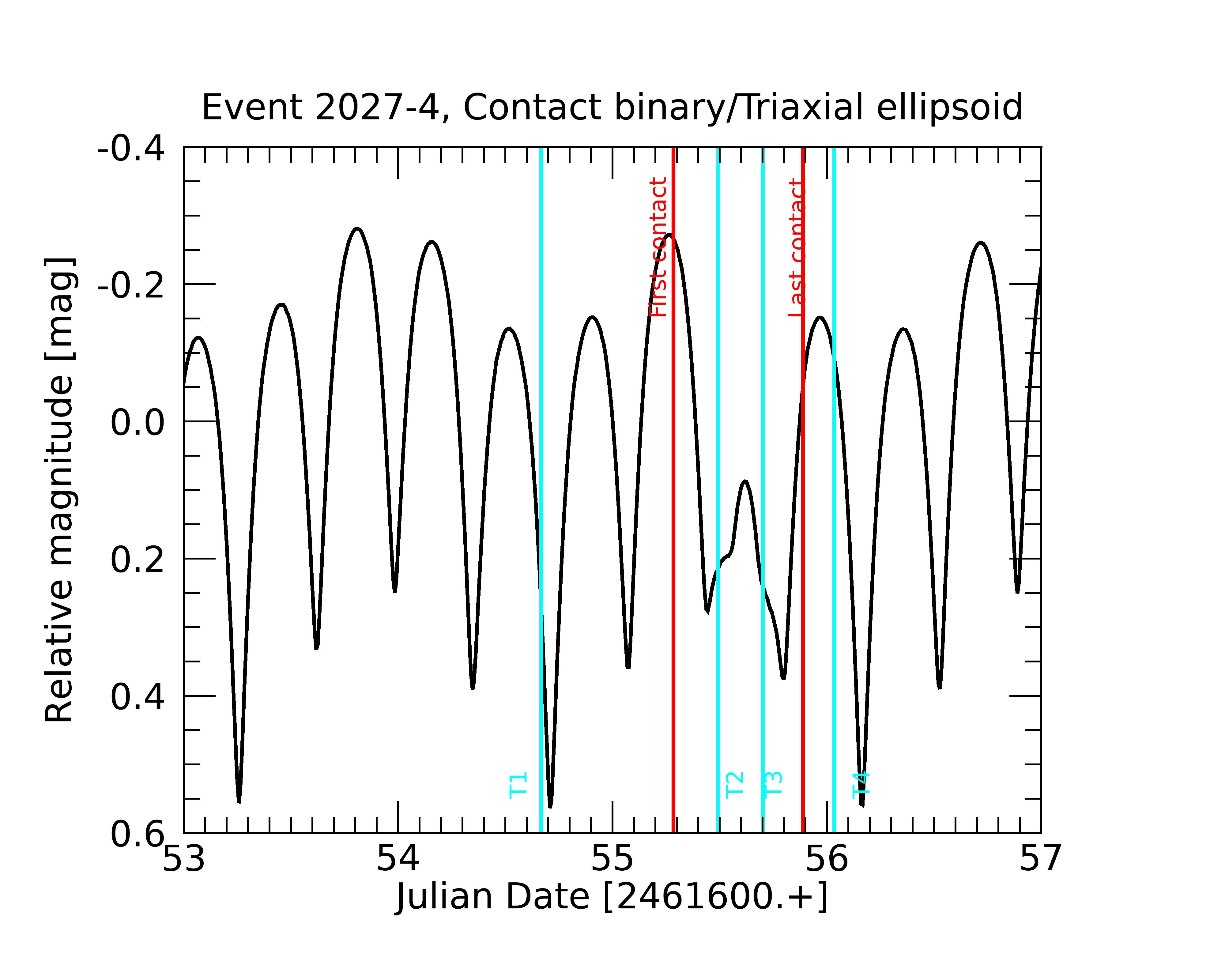}
  \caption{Modeling of the 2027-4 event assuming that Zoe is a triaxial ellipsoid.}
\label{fig:event2027-4-jacobi}
\end{figure*}


  \begin{figure*}[ht!]
\center
  \includegraphics[width=18cm, angle=0]{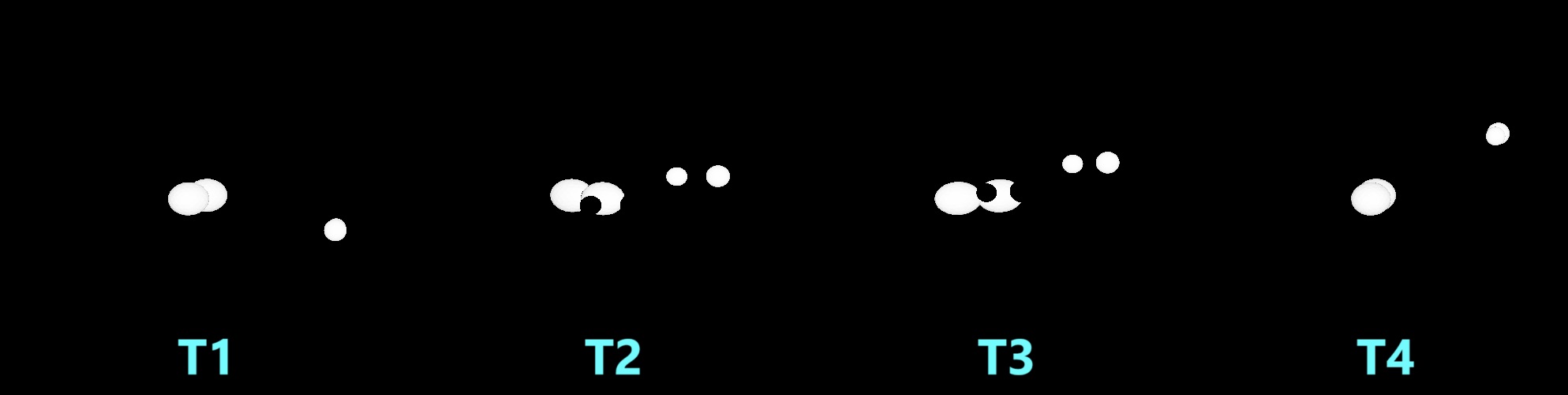}
   \includegraphics[width=10cm, angle=0]{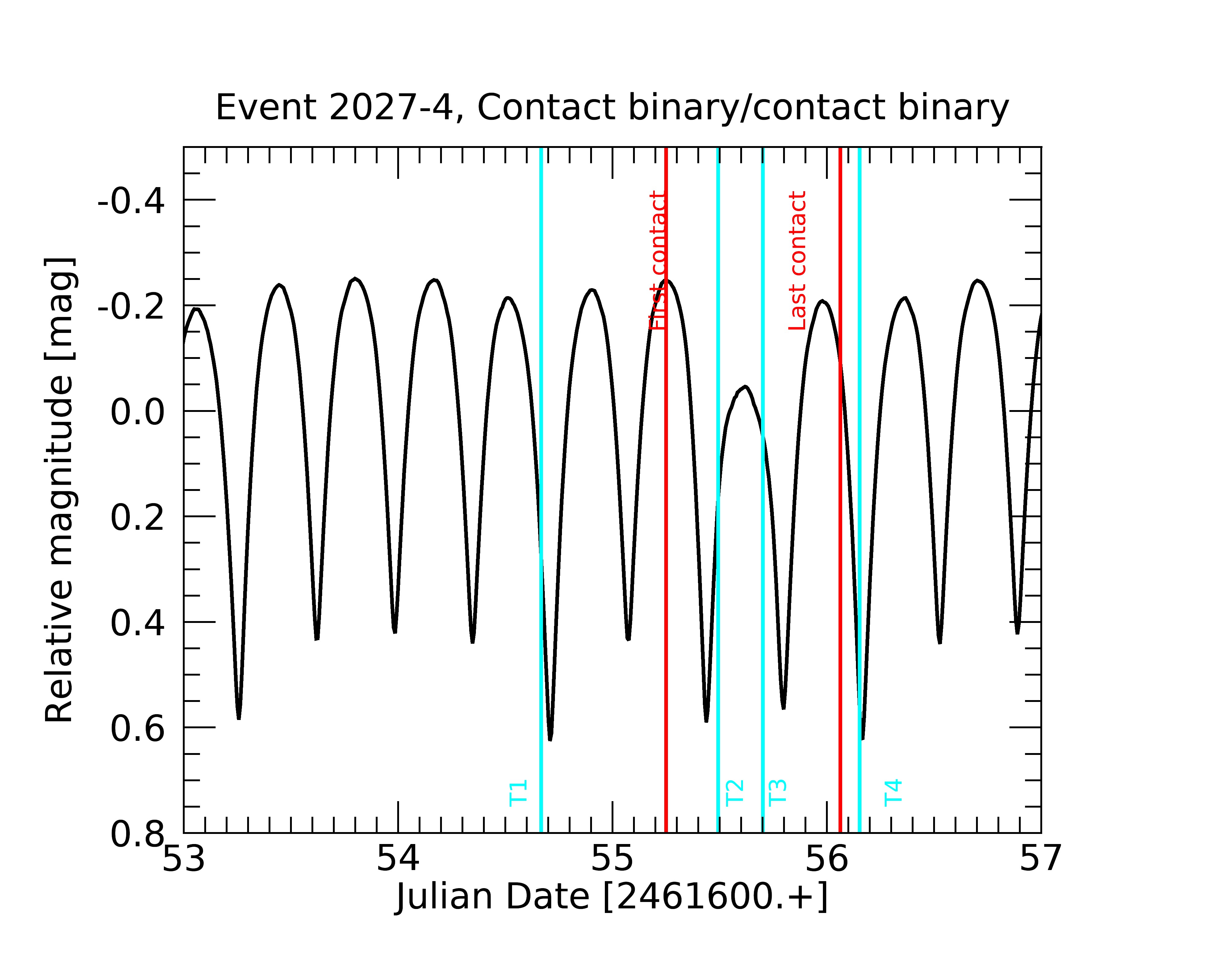}
  \caption{Modeling of the 2027-4 event assuming that Zoe is a contact binary.}
\label{fig:event2027-4-roche}
\end{figure*}

 

  \begin{figure*}[ht!]
\center
  \includegraphics[width=18cm, angle=0]{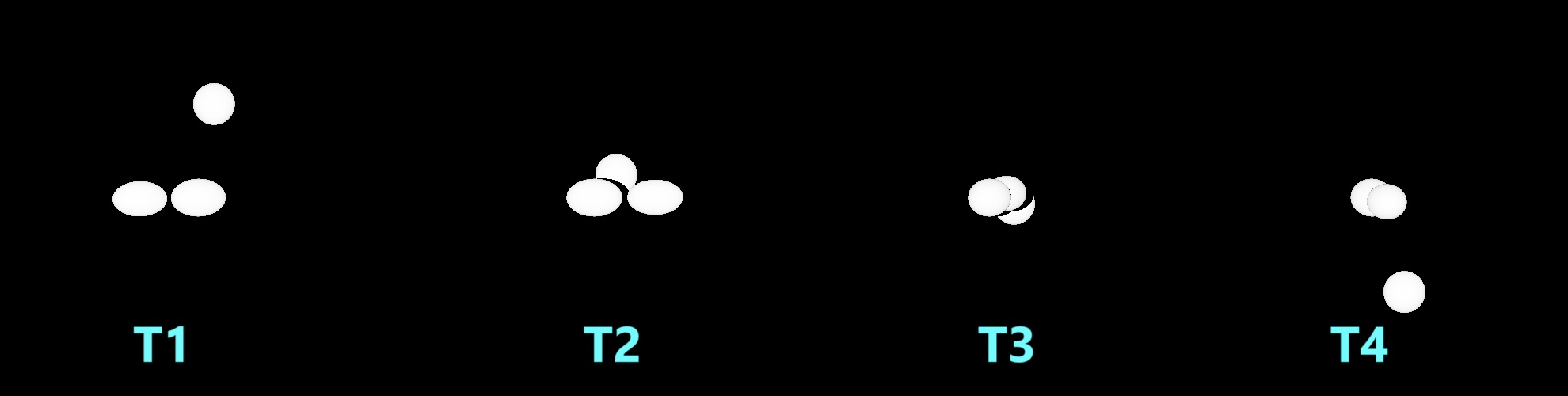}
  \includegraphics[width=10cm, angle=0]{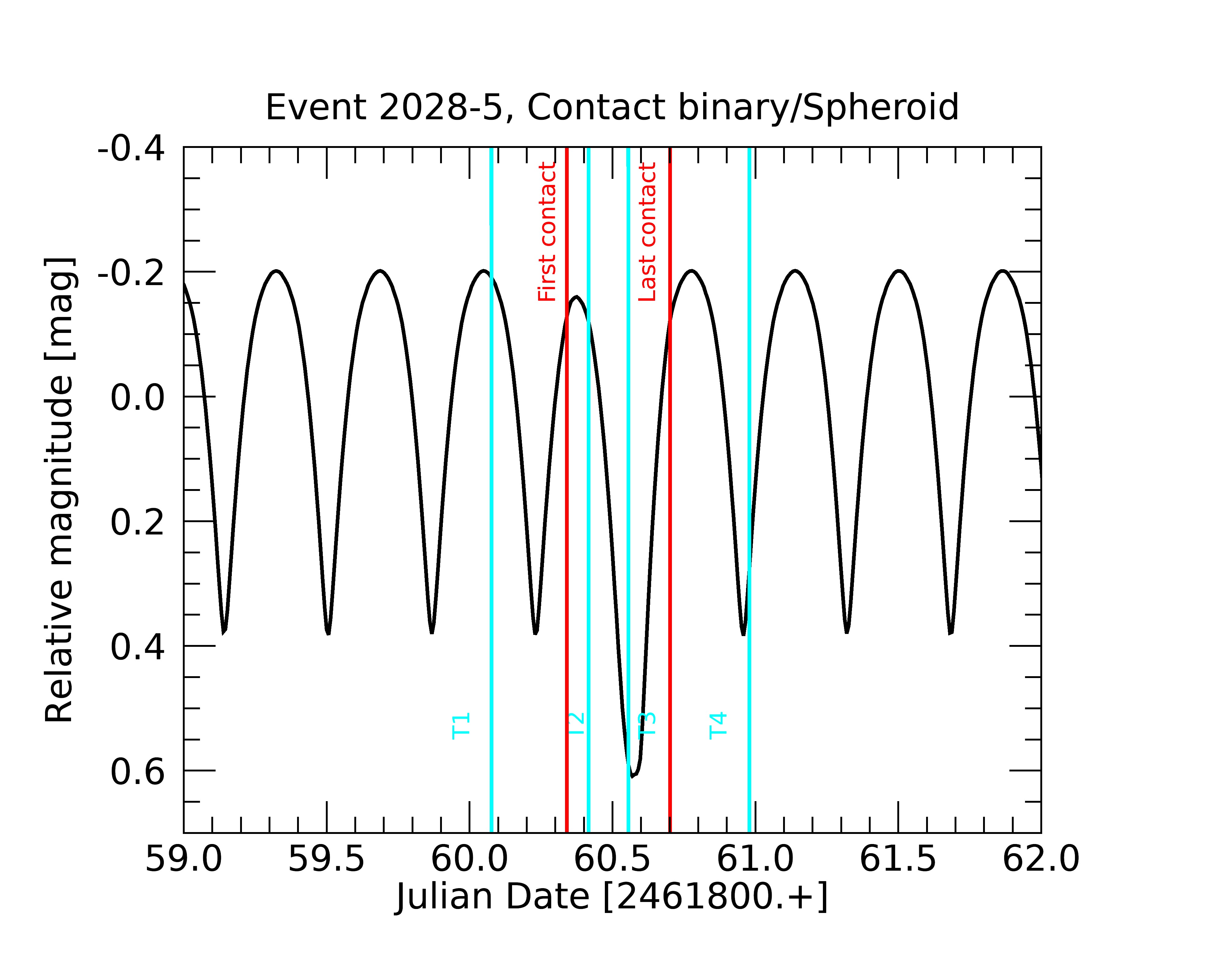}
  \caption{Modeling of the 2028-5 event assuming that Zoe is a spherical object.}
\label{fig:event2028-5-sphere}
\end{figure*}


  \begin{figure*}[ht!]
\center
  \includegraphics[width=18cm, angle=0]{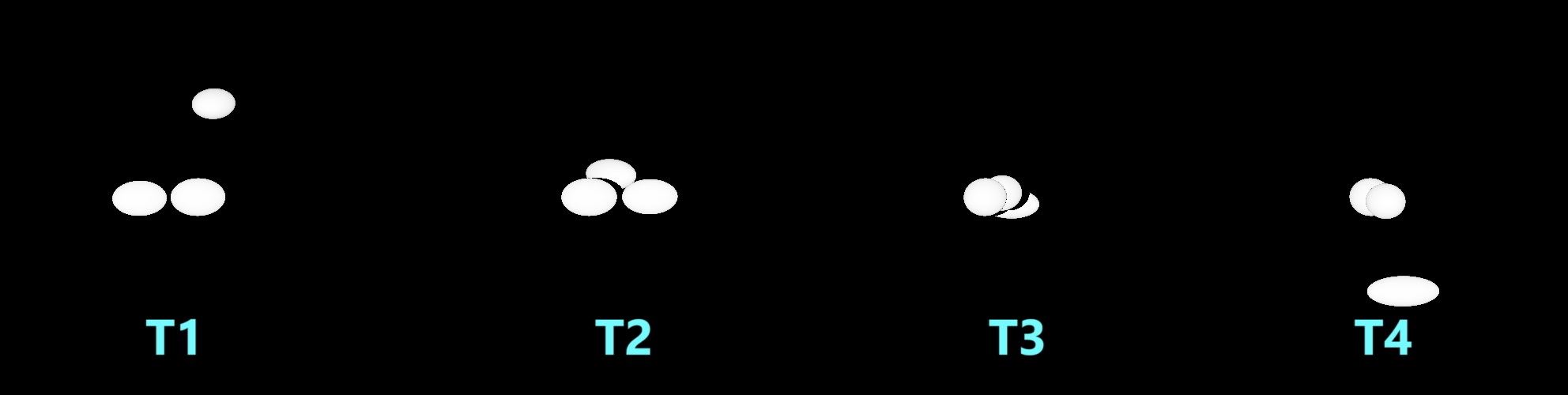}
  \includegraphics[width=10cm, angle=0]{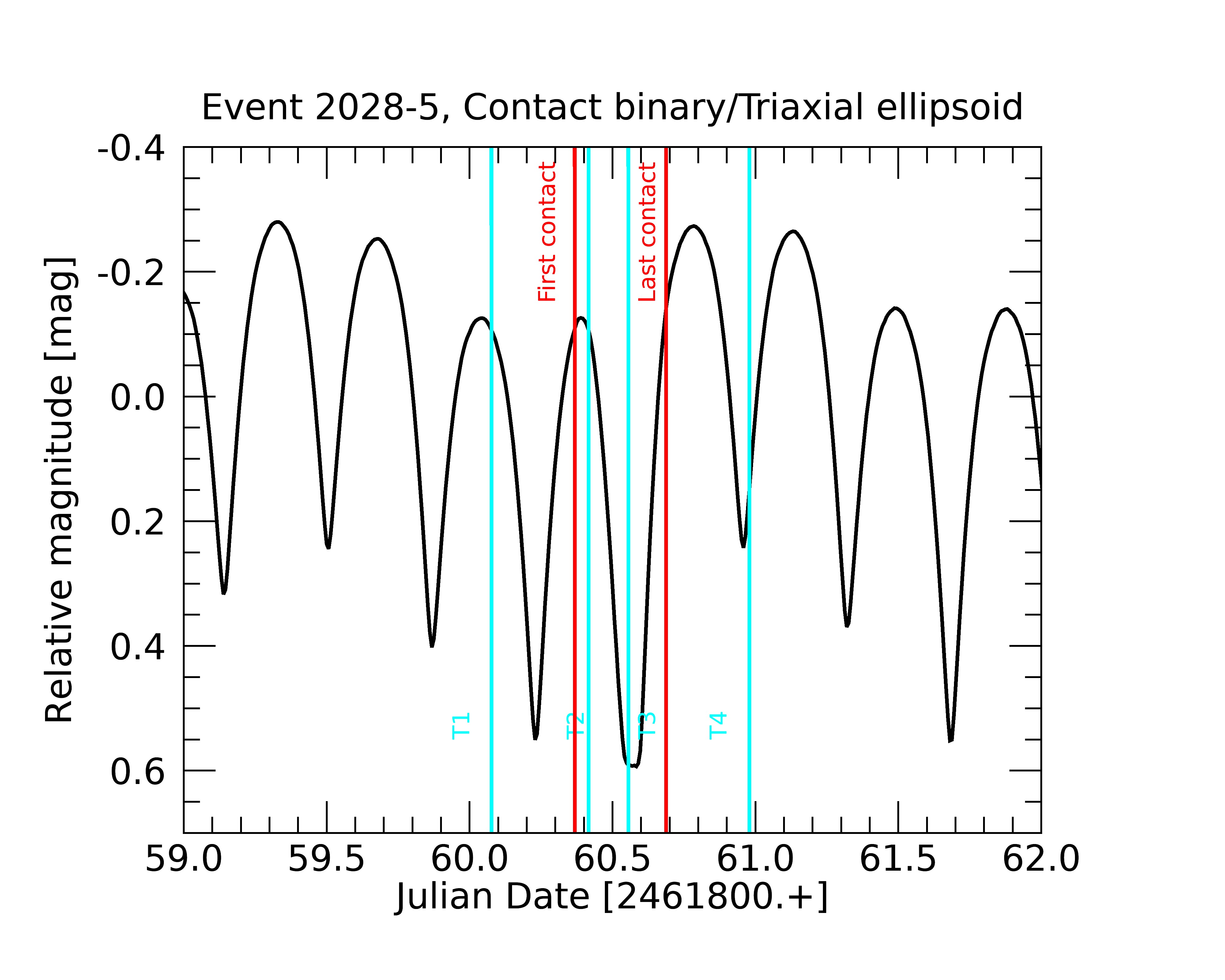}
  \caption{Modeling of the 2028-5 event assuming that Zoe is an ellipsoidal object.}
\label{fig:event2028-5-jacobi}
\end{figure*}


  \begin{figure*}[ht!]
\center
  \includegraphics[width=18cm, angle=0]{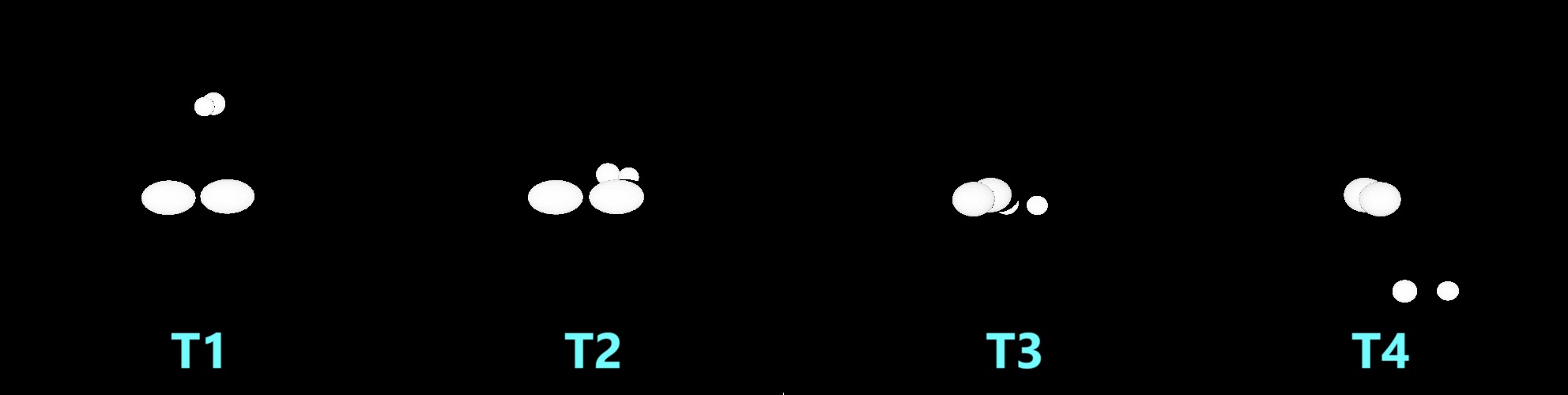}
  \includegraphics[width=10cm, angle=0]{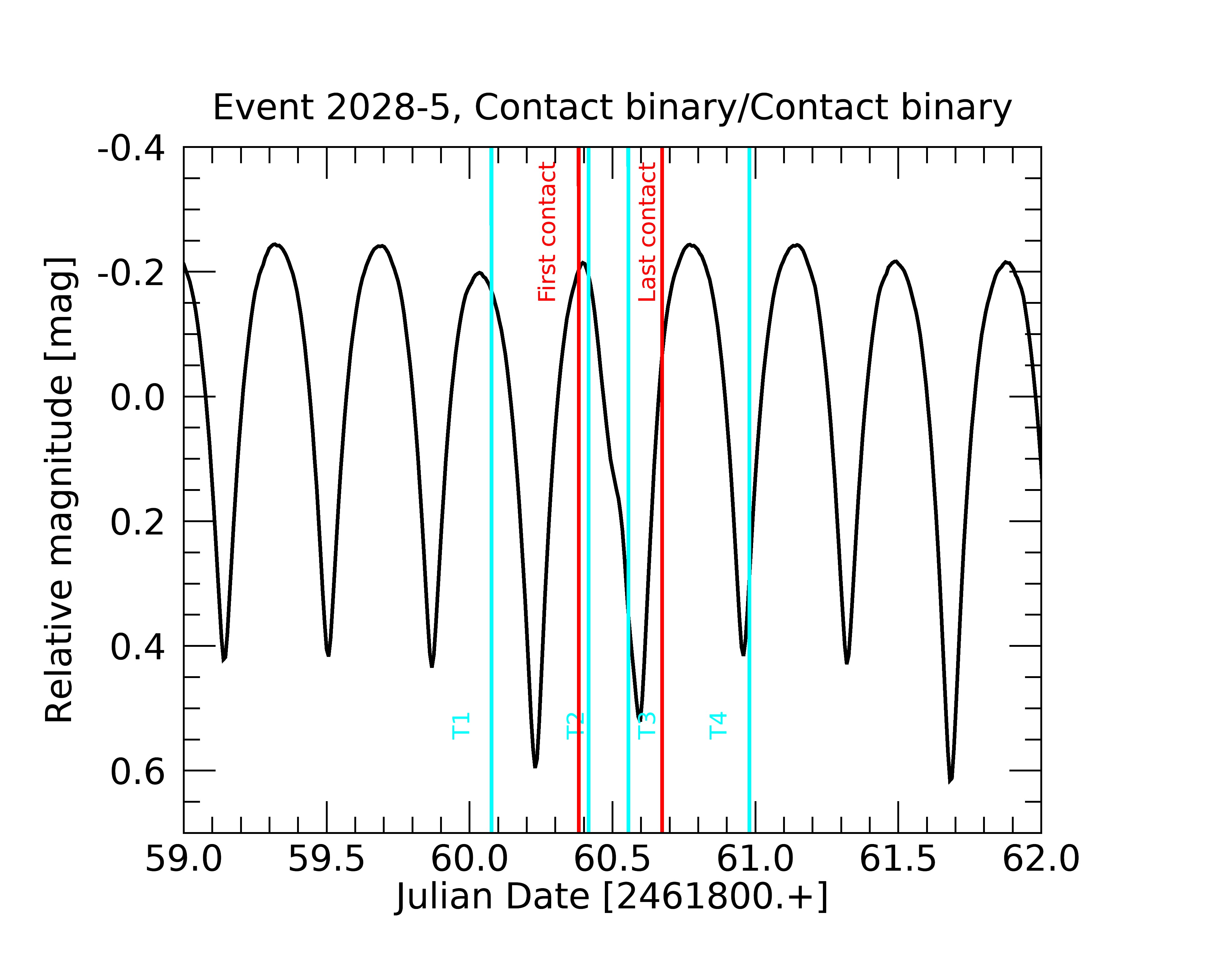}
  \caption{Modeling of the 2028-5 event assuming that Zoe is a contact binary.}
\label{fig:event2028-5-roche}
\end{figure*}


  \begin{figure*}[ht!]
\center
  \includegraphics[width=18cm, angle=0]{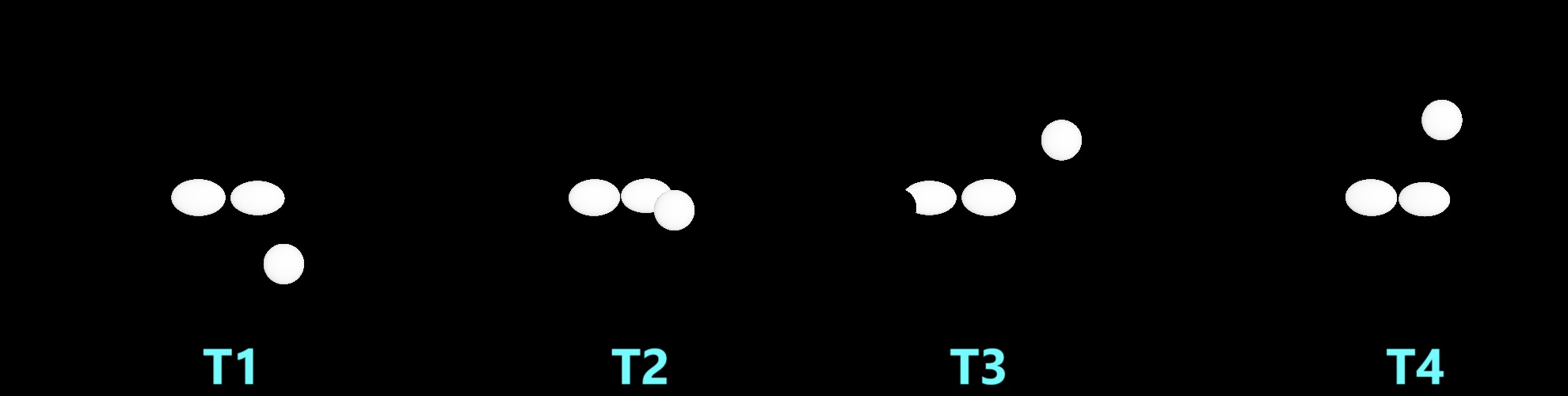}
  \includegraphics[width=10cm, angle=0]{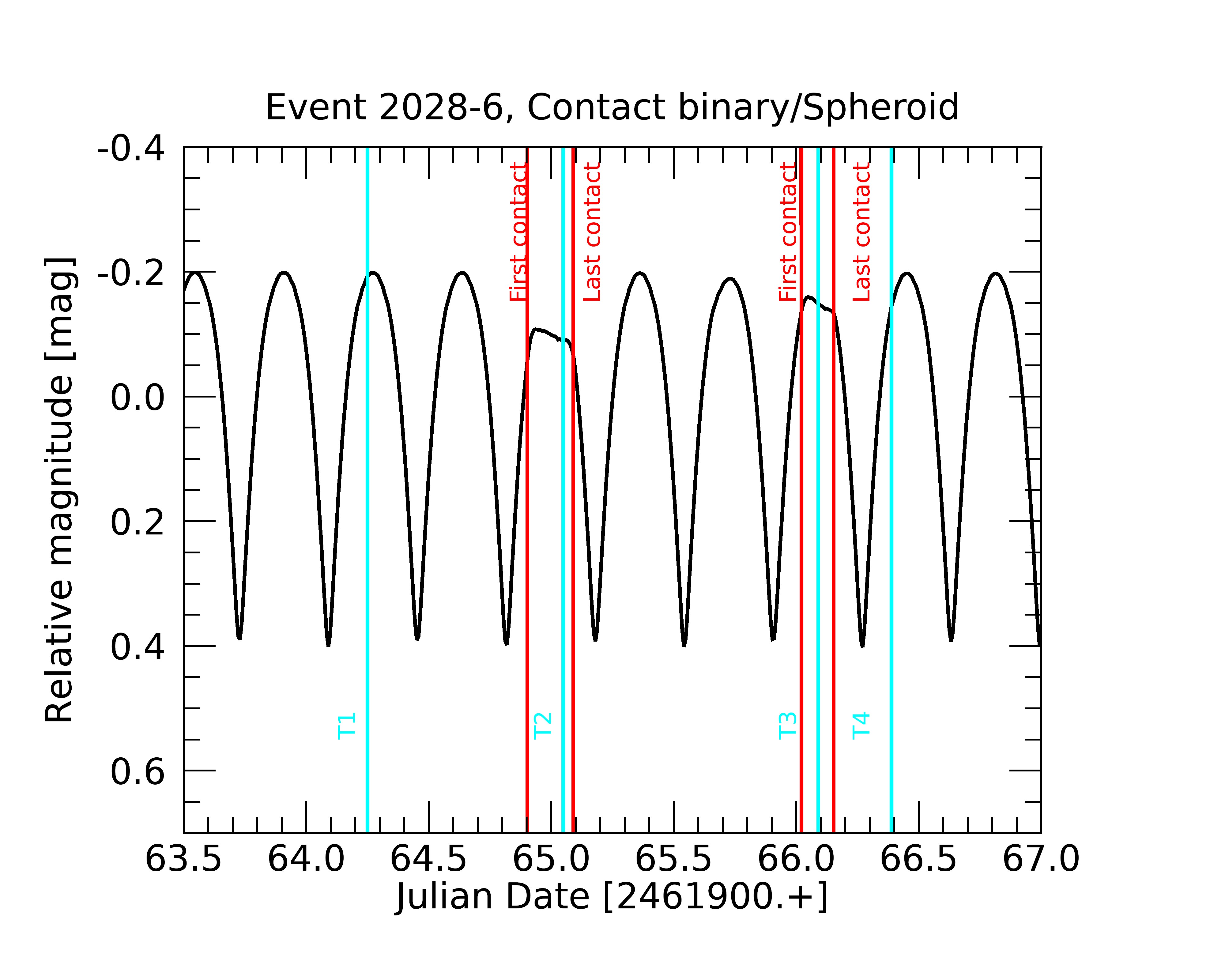}
  \caption{Modeling of the 2028-6 event assuming that Zoe is a spherical object.}
\label{fig:event2028-6-sphere}
\end{figure*} 


  \begin{figure*}[ht!]
\center
  \includegraphics[width=18cm, angle=0]{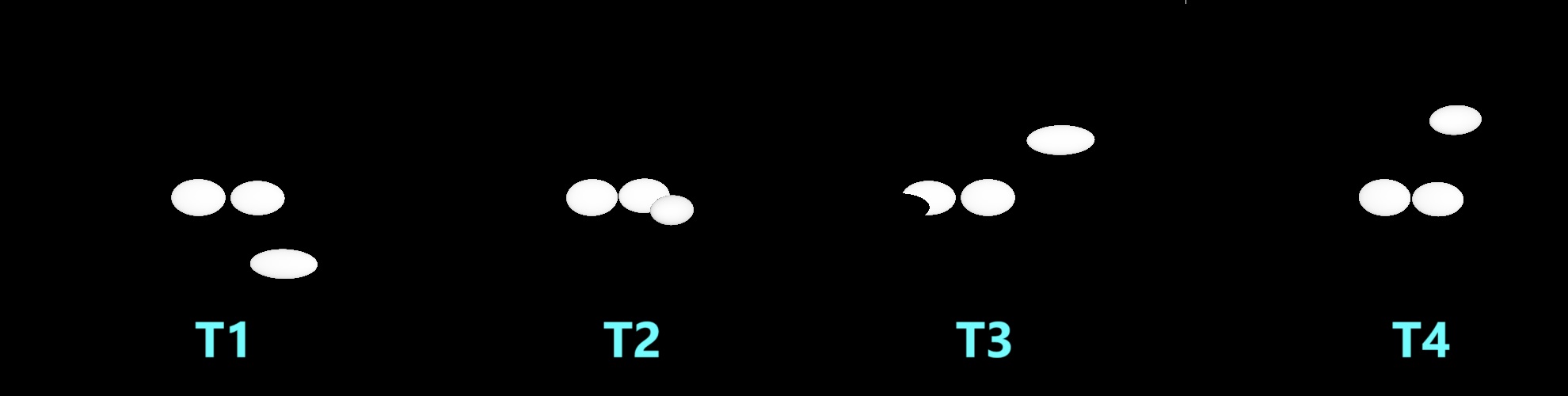}
   \includegraphics[width=10cm, angle=0]{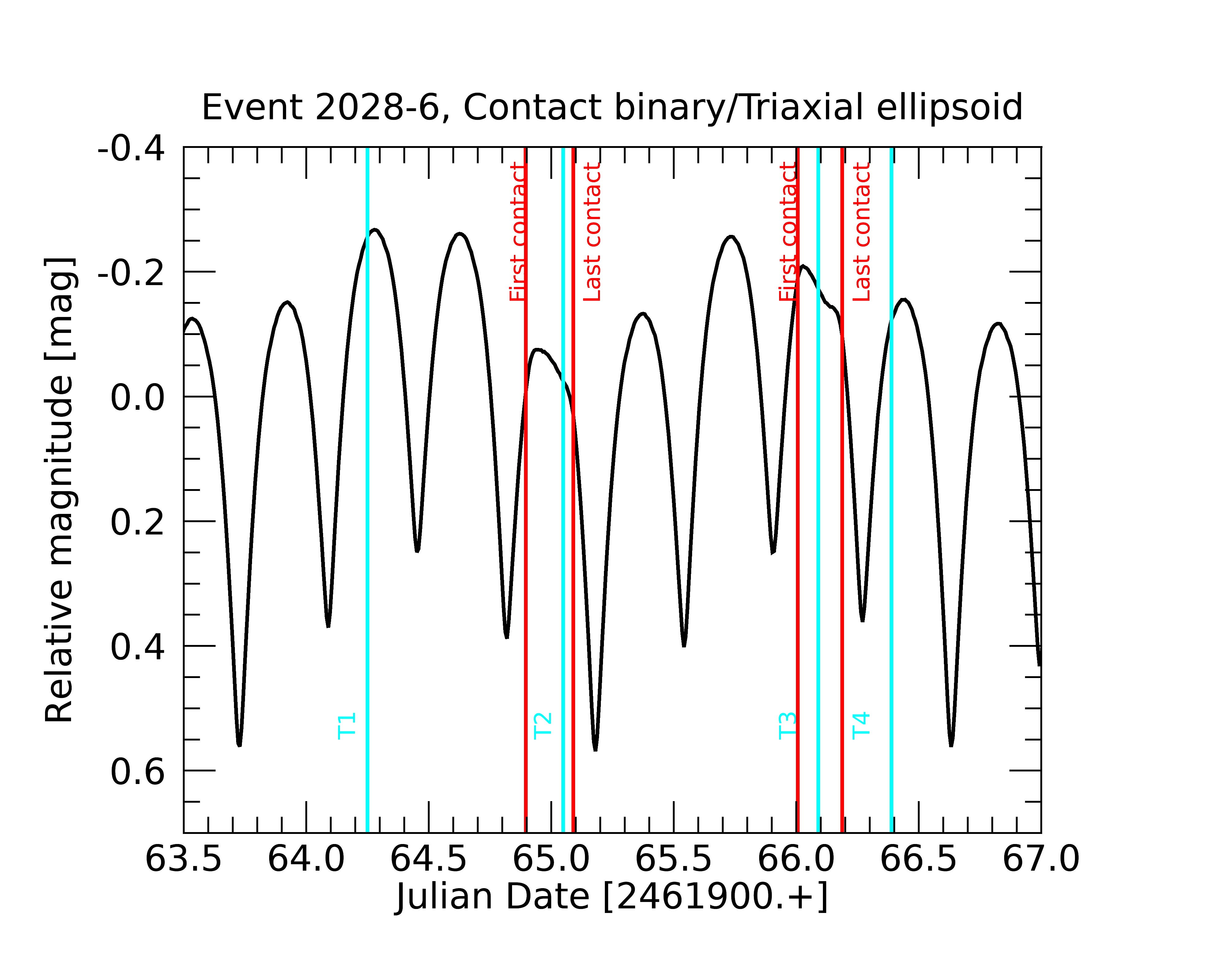}
  \caption{Modeling of the 2028-6 event assuming that Zoe is an ellipsoidal object.}
\label{fig:event2028-6-jacobi}
\end{figure*} 


  \begin{figure*}[ht!]
\center
  \includegraphics[width=18cm, angle=0]{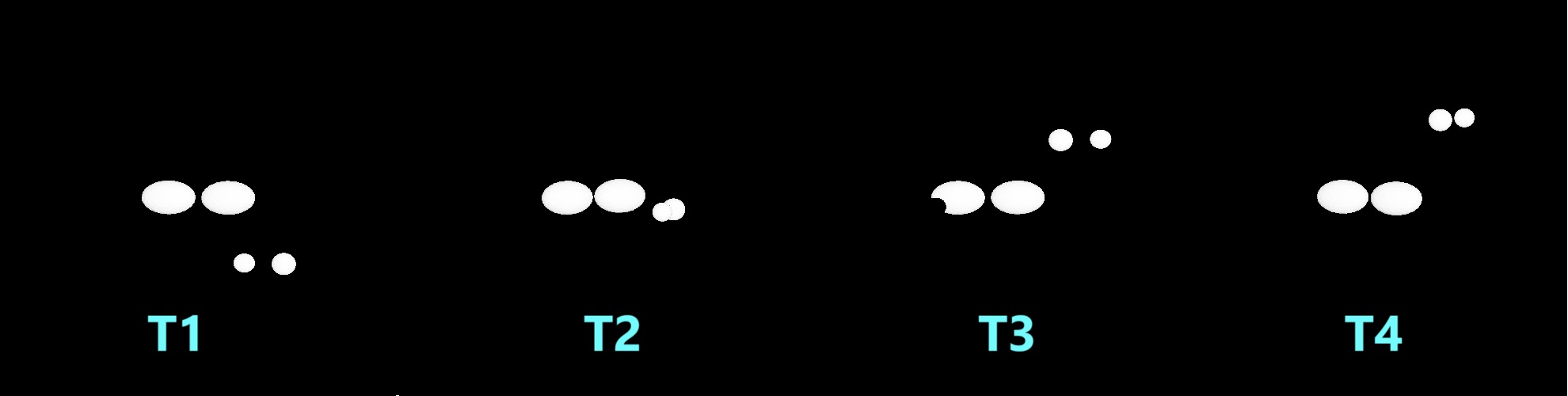}
    \includegraphics[width=10cm, angle=0]{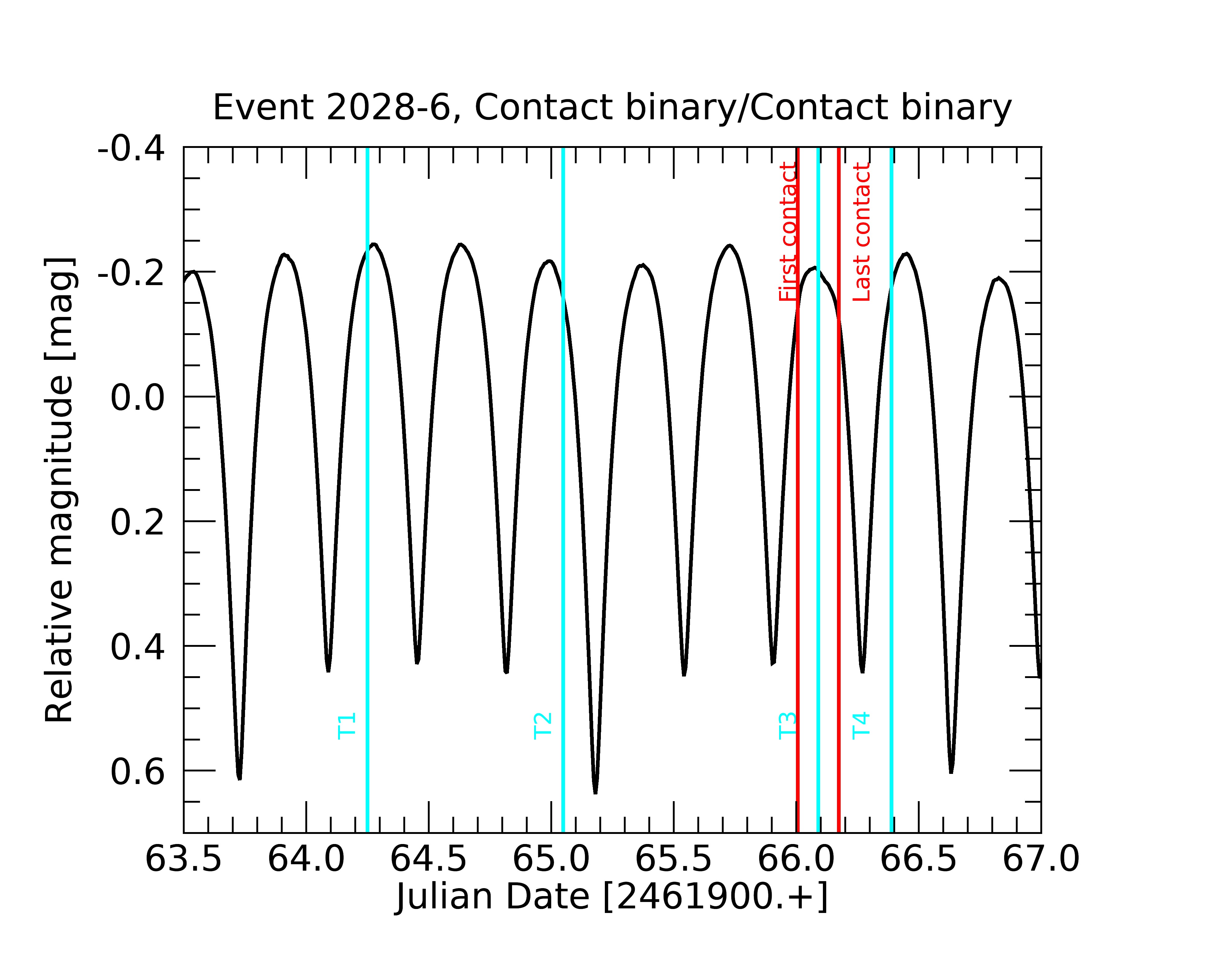}
  \caption{Modeling of the 2028-6 event assuming that Zoe is a contact binary.}
\label{fig:event2028-6-roche}
\end{figure*}


  \begin{figure*}[ht!]
\center
  \includegraphics[width=18cm, angle=0]{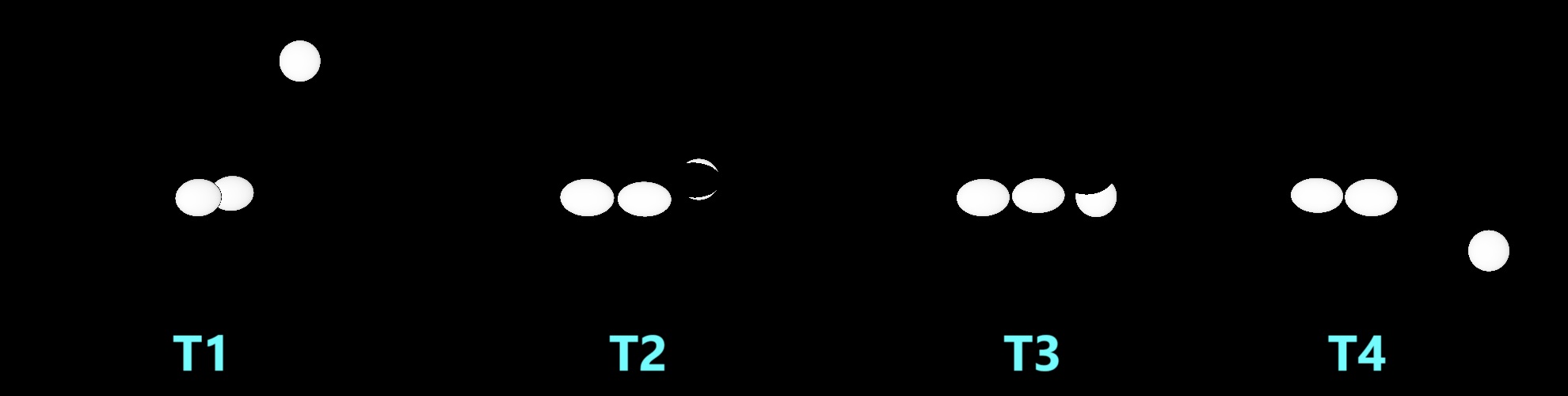}
  \includegraphics[width=10cm, angle=0]{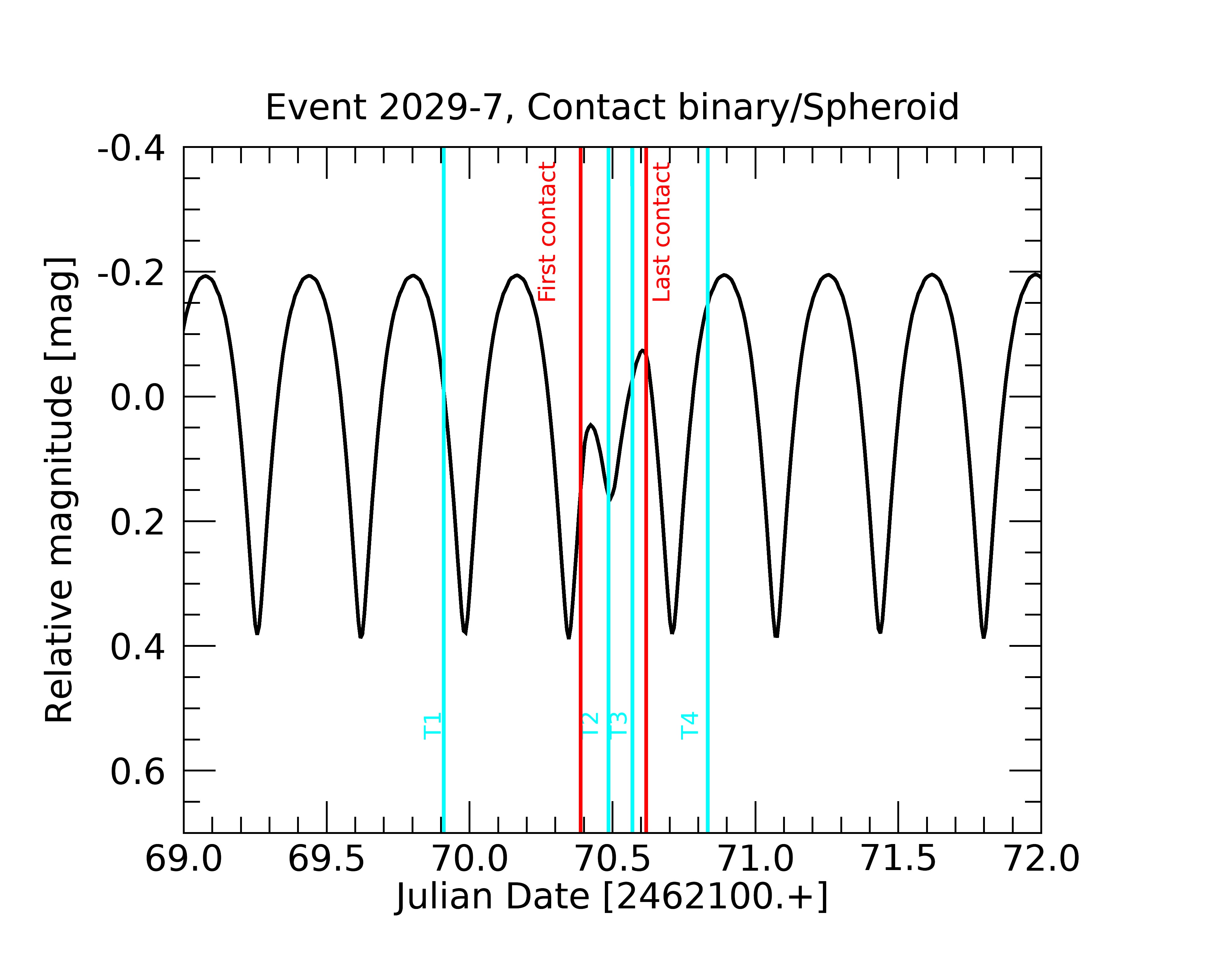}
  \caption{Modeling of the 2029-7 event assuming that Zoe is a sphere.}
\label{fig:event2029-7-sphere}
\end{figure*}


  \begin{figure*}[ht!]
\center
  \includegraphics[width=18cm, angle=0]{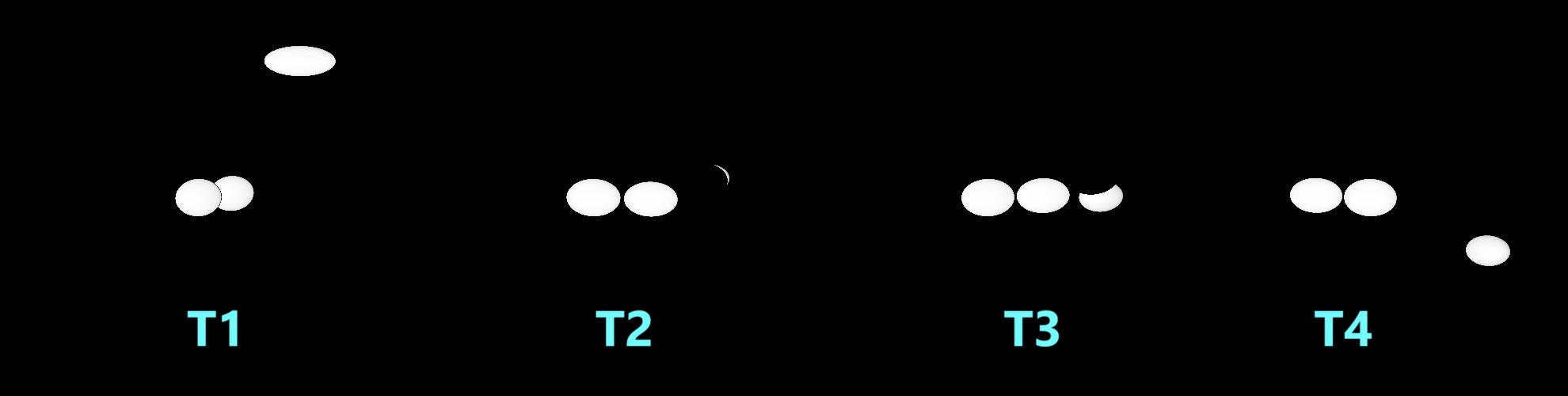}
   \includegraphics[width=10cm, angle=0]{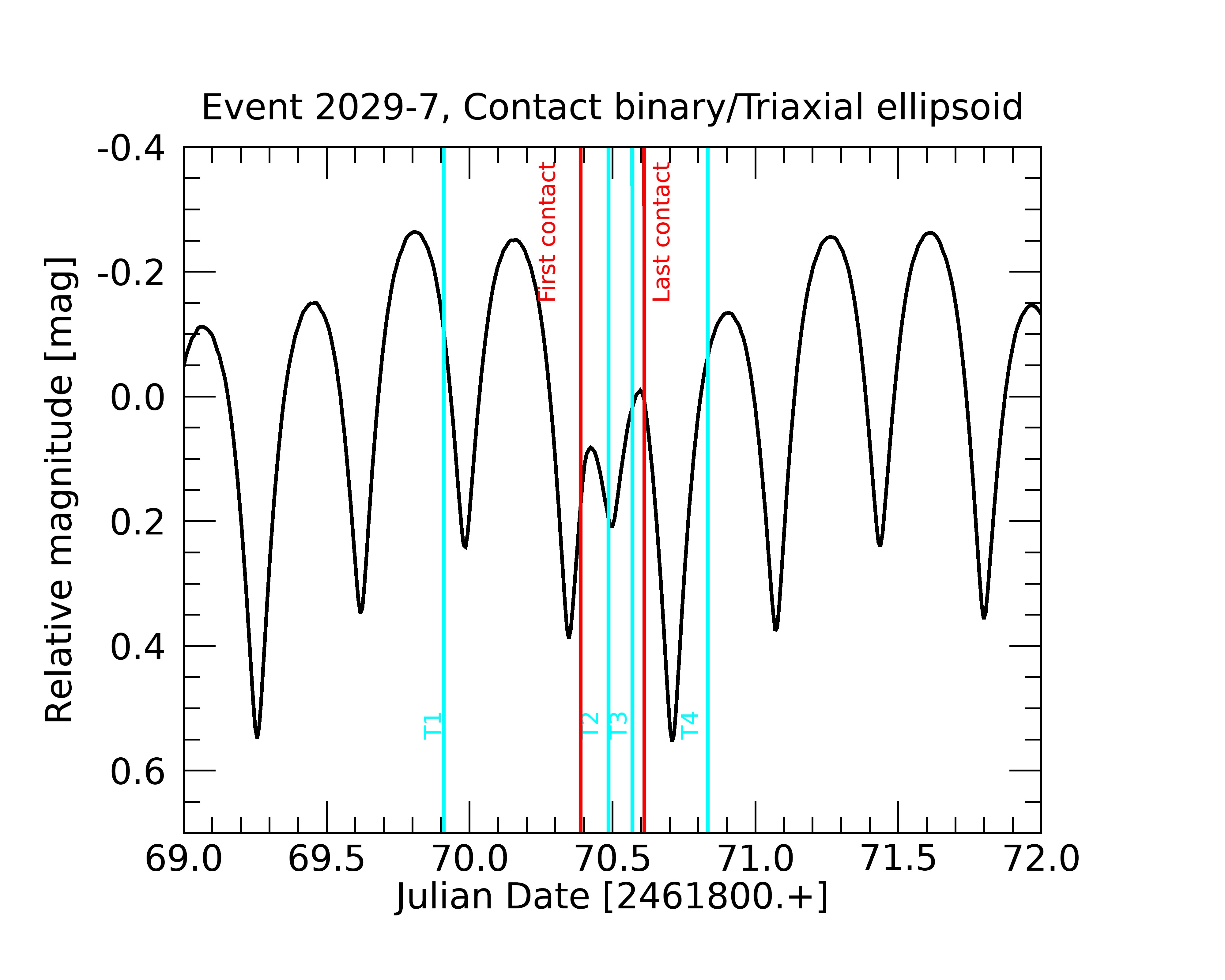}
  \caption{Modeling of the 2029-7 event assuming that Zoe is an ellipsoidal object.}
\label{fig:event2029-7-jacobi}
\end{figure*}


  \begin{figure*}[ht!]
\center
  \includegraphics[width=18cm, angle=0]{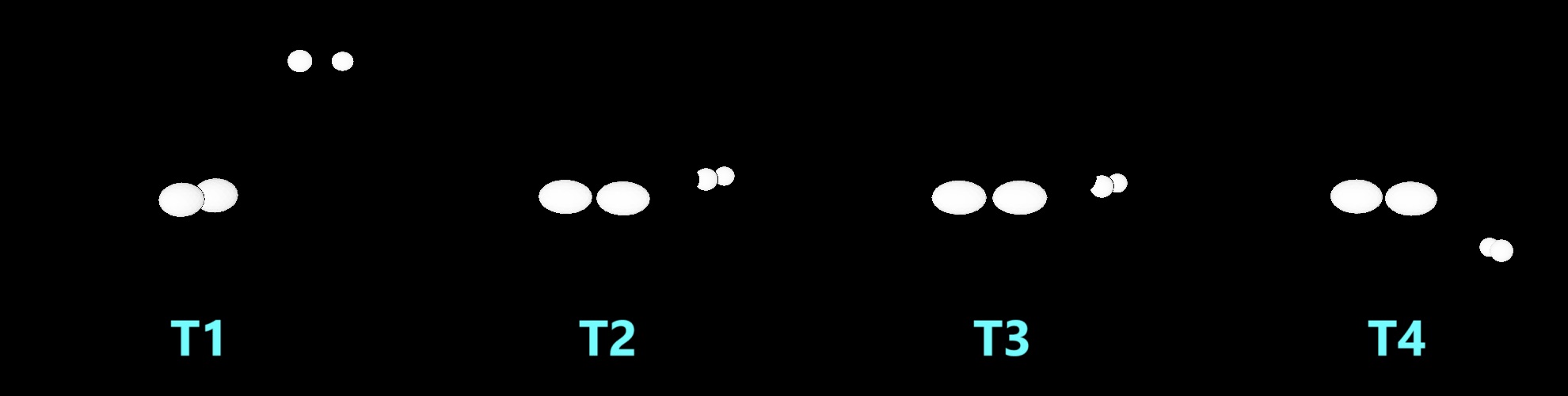}
   \includegraphics[width=10cm, angle=0]{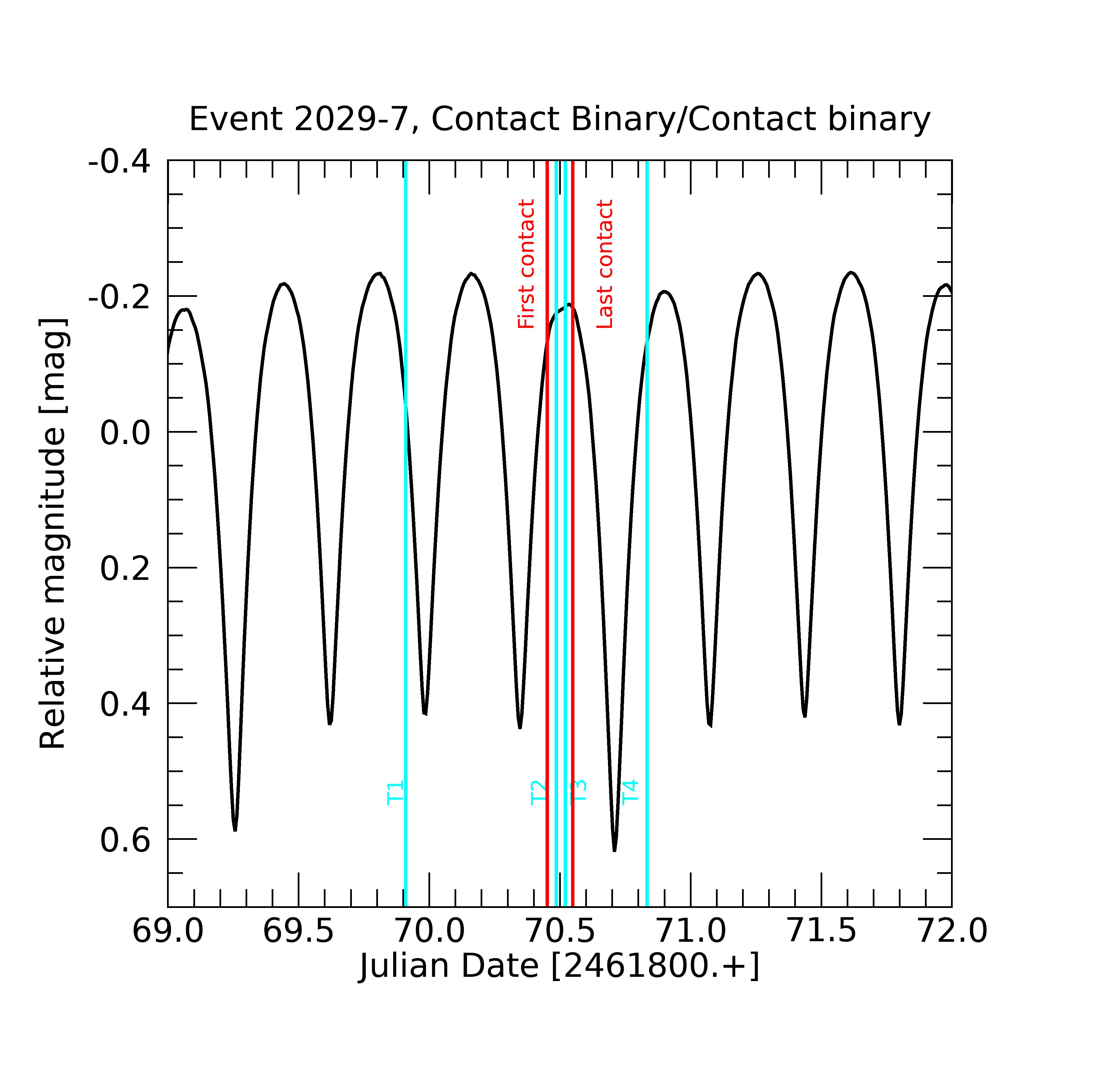}
  \caption{Modeling of the 2029-7 event assuming that Zoe is a contact binary.}
\label{fig:event2029-7-roche}
\end{figure*}


\end{document}